\documentclass[preprint2]{emulateapj}
\usepackage{graphicx, amsmath}
\usepackage{bm}

\newcommand{\EQ}{\begin{equation}}
\newcommand{\EN}{\end{equation}}
\newcommand{\EQA}{\begin{eqnarray}}
\newcommand{\ENA}{\end{eqnarray}}

\newcommand{\Eq}[1]{Equation~(\ref{#1})}
\newcommand{\Eqs}[2]{Equations~(\ref{#1}) and~(\ref{#2})}

\newcommand{\Fig}[1]{Figure~\ref{#1}}

\newcommand{\Tab}[1]{Table~\ref{#1}}

{}
{}
{}

{}
{}
{}
{}
{}
{}
{}
{}
{}
{}
{}
{}
{}
{}
{}
{}
{}
{}
{}
{}

{}
{}
{}

{}

{}
{}

\newcommand{\yyy}{\hat{\mbox{\boldmath $y$}} {}}

\newcommand{\yy}{\mbox{\boldmath $y$} {}}
\newcommand{\zz}{\mbox{\boldmath $z$} {}}
\newcommand{\vv}{\mbox{\boldmath $v$} {}}

\newcommand{\kk}{\bm{k}}

\newcommand{\xx}{\bm{x}}

\newcommand{\BB}{\mbox{\boldmath $B$} {}}

\newcommand{\JJ}{\mbox{\boldmath $J$} {}}

\newcommand{\nab}{\mbox{\boldmath $\nabla$} {}}

\def\Rm{\mbox{\rm Re}_M}

\newcommand{\J}{\,{\rm J}}

\newcommand{\ov}[1]{\overline{ #1}}

\begin{document}

\title{Temperature Fluctuations driven by  Magnetorotational Instability in Protoplanetary Disks}
\author{Colin P. McNally\altaffilmark{1}, Alexander Hubbard\altaffilmark{2}, Chao-Chin Yang\altaffilmark{3}, and Mordecai-Mark
  Mac Low\altaffilmark{2}}
\altaffiltext{1}{Niels Bohr International Academy, Niels Bohr Institute, Blegdamsvej 17, DK-2100, Copenhagen~\O, Denmark,
\email{cmcnally@nbi.dk}}
\altaffiltext{2}{Department of Astrophysics, American Museum of Natural History, New York, NY 10024-5192, USA
\email{ahubbard@amnh.org,mordecai@amnh.org}
}
\altaffiltext{3}{Lund Observatory, Department of Astronomy and
  Theoretical Physics, Lund University, Box 43, SE-22100, Lund, Sweden
\email{ccyang@astro.lu.se}
}

\begin{abstract}

  The magnetorotational instability (MRI) drives magnetized turbulence in
  sufficiently ionized regions of protoplanetary disks, leading to
  mass accretion.  The dissipation of the potential energy associated
  with this accretion determines the thermal structure of accreting
  regions. Until recently, the heating from the turbulence has only
  been treated in an azimuthally averaged sense, neglecting local
  fluctuations.  However, magnetized turbulence dissipates its energy
  intermittently in current sheet structures. We study this
  intermittent energy dissipation using high resolution numerical
  models including a treatment of radiative thermal diffusion in an
  optically thick regime.  Our models predict that these turbulent
  current sheets drive order-unity temperature variations even where
  the MRI is damped strongly by Ohmic resistivity.  This implies that
  the current sheet structures where energy dissipation occurs must be
  well resolved to correctly capture the flow structure in numerical
  models.  Higher resolutions are required to resolve energy
  dissipation than to resolve the magnetic field strength or accretion
  stresses.  The temperature variations are large enough to have major
  consequences for mineral formation in disks, including melting
  chondrules, remelting calcium-aluminum rich inclusions, and
  annealing silicates; and may drive hysteresis:
  current sheets in MRI active regions could be significantly more
  conductive than the remainder of the disk.

\end{abstract}

\section{Introduction}
In regions of accretion disks where the magnetorotational 
instability (MRI) acts, differential rotation
shears magnetic fields, producing turbulence.  The resulting torques
extract gravitational potential energy and drive 
accretion flows \citep{1998RvMP...70....1B}.  In
a steady state, the extracted energy must be either exported in a
wind, or dissipated locally, heating the disk, which then cools radiatively.
In the case of local dissipation, the strength of the accretion
flow depends on the nature of the dissipation. The exact nature of the
dissipation may also determine the course of mineral formation in
protoplanetary disks, as the total amount of energy dissipated suffices
to thermally process the solids and ices present
\citep{2010MNRAS.404.1903K}. 
The meteoritic record, particularly the chondrites, may reflect these processes.

If that energy dissipation were evenly distributed, it would have little
effect on local temperatures.  This forms the basis for the common
approximation of local isothermality in MRI simulations.
However, volume averaged quantities can mislead because the MRI
amplified magnetic field, and all its dependent quantities, vary
significantly in both time and space.
In particular, magnetized
turbulence dissipates its energy in current sheets
\citep{1972ApJ...174..499P,1994ISAA....1.....P,1997PhR...283..227C},
quasi-2D structures where the magnetic fields change rapidly in space.
This inhomogeneity in space and time is interesting not just because
it may control the strength of the accretion flow, but also because it
must lead to concentration of energy dissipation, and thus spatial
variations in temperature.  Indeed, \citet{hirose11} found that even
in regions where stellar irradiation dominates the energy budget,
current sheets can locally heat gas to temperatures $50\%$ greater
than that of the gas heated by starlight.

Most simulations of the MRI in
protoplanetary disks are at least locally isothermal in design: they do not advance a
temperature equation, although the imposed temperature may be a
function of radial position.
Some of the earliest MRI simulations did include 
an energy equation, Ohmic heating, thermal diffusion, and a balancing cooling, but
these were only low resolution and the Ohmic heating was not resolved
\citep[e.g.,][]{1995ApJ...446..741B,1996ApJ...458L..45B}.

More recently, some simulations have solved
the full radiative transfer problem to determine the temperature
structure of a local region of the disk
\citep{turner03,turner04,hirose06,blaes07,krolik07,flaig09}, and even
included stellar irradiation of the disk surface in the case of
\citet{hirose11}.  A first non-isothermal, global model
including radiative transfer was performed by
\citet{2013A&A...560A..43F}, assuming an initial azimuthal magnetic
field geometry.  However, none of these models, except those of \citet{hirose11}, included an explicit
treatment of heating and magnetic field diffusion caused by resistivity. 
\citet{hirose11} included
a detailed model of the resistivity in order to study the extent
of the MRI-dead zone, but their simulations 
appear too poorly resolved to capture the full structure of current sheet driven heating.
The current sheets they studied form in the upper active layers of a disk section with a midplane dead zone.
These current sheets develop where large azimuthally directed flux tubes are driven together.
This basic behavior, of the strongest current sheets occurring where oppositely directed 
azimuthal flux tubes contact, is also found in the simulations we
describe here.

\citet{2000ApJ...530..464F} studied net vertical field MRI with finite Ohmic resistivity for a range of resistivities.
Two particular qualitative features of those models recur in our study.
First, the variation of the Maxwell stress is found to be much greater
in MRI with significant Ohmic resistivity than is found at low resistivity, 
and this variation appears to be  connected to quasi-periodic reappearance of MRI channel flows.
Second, considerable heating though Ohmic dissipation occurs in their more resistive models. 
However, their models lack an energy loss mechanism that would allow the system to reach a quasi-steady 
state at long times. 
The model we study here includes such a mechanism.

Characterizing the heating effects in current sheets is particularly
important because in thermally ionized regions of protoplanetary
disks, spatial temperature variation can drive short-circuit
instabilities \citep{2012ApJ...761...58H,2013ApJ...767L...2M}.  This
occurs because the
ionization fraction, and hence the Ohmic resistivity, is an
exquisitely sensitive function of temperature when the temperature is
high enough for thermal ionization to set in.  
If a current sheet forms, Ohmic heating can raise the temperature,
increasing the ionization level, which reduces the
resistivity, concentrating the current.
This leads to even stronger Ohmic heating, causing runaway heating
in the sheet. 

Even in regions where non-thermal ionization dominates, order unity
temperature variations have important effects.  In hotter regions,
order-unity temperature variations can process rocks, annealing
amorphous silicates or melting chondrules. They impact the behavior of
the MRI by altering the density and pressure structure of the gas, as
well as the strength of ambipolar diffusion (ion-neutral drift) and
the Hall effect.
Furthermore, even modest temperature variations will transform
ice lines into broad regions with thickness a large fraction of their orbital
radius, allowing solids and vapor to coexist at the same radial 
position, with repeated evaporation-condensation cycles \citep{2013A&A...552A.137R}.  This could
strengthen, compactify, and soften dust grains, allowing 
for both condensation based growth and enhanced collisional growth.

In this paper we describe local models of
MRI including Ohmic resistivity, at sufficiently high resolution to explicitly resolve at least
some current sheet structure, and including the full energy equation
and an approximate treatment of radiative
transport. This allows us to study the spatial and thermal
structure of the current sheets.  We give an analytic argument that strong temperature
fluctuations should be expected in MRI-active regions under some conditions, and indeed find
such fluctuations in our models.  Further, we found
that the resolution requirements to fully capture the dissipation are
higher than generally thought, so that many existing
studies, such as \citet{hirose11}, appear under-resolved.

\begin{figure}
\includegraphics[width=\columnwidth]{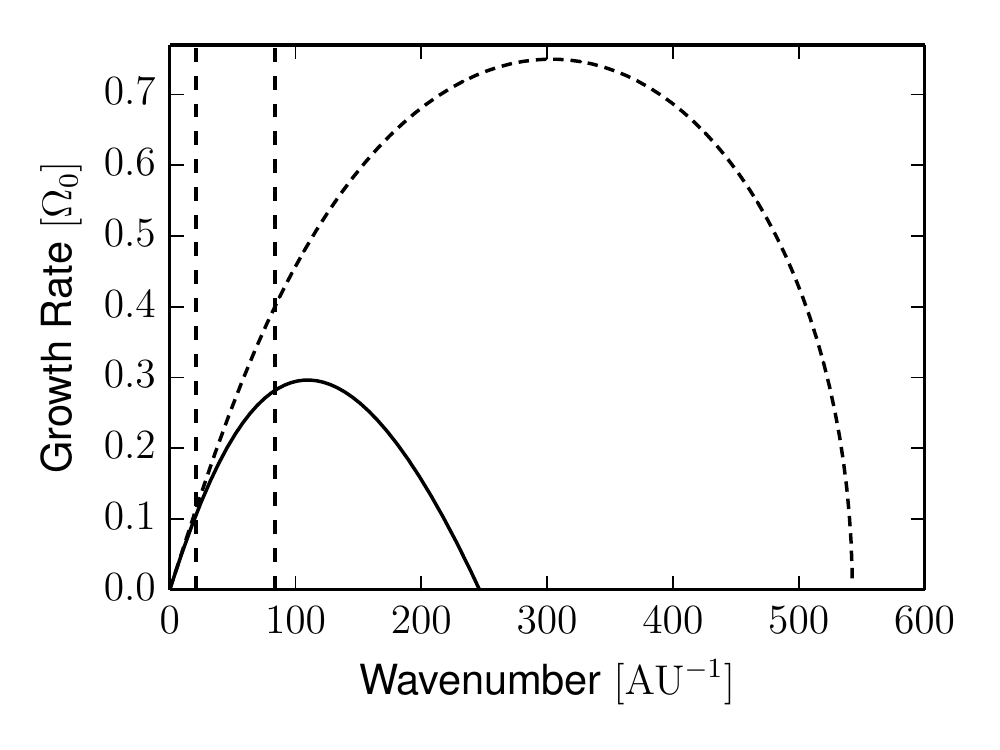}
\caption{Growth rates of zero-resistivity MRI and the specific case of resistive MRI considered here. 
{ Dashed:} zero resistivity, {Solid:} Elsasser  number $\Lambda=0.5$.
{ Vertical Long Dashed:} wavenumbers corresponding to the vertical
heights (smallest vertical wavenumber) of 
 the two sizes of shearing box domain of $H$ and $4H$ used for the simulations in this work.
}
\label{fig_dispersion}
\end{figure}

The effect of resistivity on the MRI dispersion relation can be
parameterized by the Elsasser number
\begin{align}
\Lambda \equiv \frac{v_{\rm A}^2}{\eta \Omega_0}.
\label{Elsasser}
\end{align}
In the literature discussing Ohmic resistivity in the context of the
MRI, this quantity has also been termed the magnetic
Reynolds number or the Lundquist number.  In
\Fig{fig_dispersion} we show the dispersion relation of the MRI for
the parameters studied in this paper (see \Tab{table_parameters}) as
well as for the ideal case 
\citep{1996ApJ...457..798J,1999ApJ...515..776S,
  2008ApJ...684..498P}.  
Considering incompressible axisymmetric perturbations with a net vertical field  and Ohmic resistivity,
the dispersion relation has two important regimes.
For the low resistivity regime, where the Elsasser number is $\Lambda \gg 1$
the most unstable MRI wavelength increases with increasing magnetic field strength
and the fastest growing mode growth rate is  $\sim 0.75 \Omega_0$.
However, for the resistive regime with  Elsasser numbers $\Lambda < 1$, the most
unstable MRI wavelength instead decreases with increasing field strength
while the fastest growing mode growth rate decreases.
 The physical regime studied in this paper falls within this second case.

In Section~\ref{methods} we describe our numerical methods and initial
conditions.  In Section~\ref{sec_results} we present our numerical results on
the structure and evolution of current sheets formed in our
simulations, and discuss their reliability in
Section~\ref{sec_Resolution}. Section~\ref{sec_theory} gives an analytic
explanation of these results. Section~\ref{opacity} explores how
our results depend on the behavior of the opacity and the resistivity,
respectively. Finally, in Section~\ref{conclusions} we discuss the
limitations and implications of our work.

\section{Simulations}
\label{methods}

\subsection{Methods}

We performed simulations of unstratified, magnetized, shearing boxes
with net vertical field including Ohmic resistivity and radiative
thermal diffusion using the Pencil Code\footnote{Details on
  the Pencil Code and download information can be found at
  \url{http://www.nordita.org/software/pencil-code/}.},
a sixth-order, central difference in space method,
\citep{2002CoPhC.147..471B}, as well as confirming our major results
with Athena\footnote{Details and download information for Athena can be found at
\url{https://trac.princeton.edu/Athena/}.}
\citep{2008ApJS..178..137S,2010ApJS..189..142S}, a
constrained-transport, Godunov method.  Here we use Athena configured
with orbital advection, linear reconstruction, and the HLLD
Riemann solver. 

The field variables evolved in the Pencil Code were density, velocity,
magnetic vector potential, and thermal energy density.  The Pencil
Code requires diffusion operators for stability, so we employed a
grid-scaled hyperdiffusion in velocity and density (which have no
physically resolved diffusion effect in the model), along with
shock diffusion on all fields so that all shocks are captured at the
grid scale.  Because the Pencil Code is sixth order accurate in space and third order accurate in time, it
resolves small scale subsonic flow structures almost as well as a
spectral code.  However, it does not conserve energy exactly.  To
confirm our results, we replicated the configuration in Athena, a locally energy-conservative, finite volume code.
In such a method, no explicit stabilizing diffusion on the momentum
and density fields is required.
Furthermore, the conversion of energy from kinetic and magnetic energy in shocks and current 
sheets to thermal energy is fully conservative.

We included two explicit physical diffusive effects in our model,
resistivity $\eta$, and radiative thermal diffusion. We allowed neither
$\eta$ nor the opacity $\kappa$ to
vary with temperature, unlike in an actual disk.   
The form of our 
Ohmic resistivity is
\begin{align}
\eta \equiv \frac{c^2}{4 \pi \sigma}
\label{eta}
\end{align} 
in Gaussian cgs units, where $\sigma$ is the conductivity.
The evolution of the magnetic field $\BB$ is then described by the induction equation
\begin{equation} \frac{\partial \BB}{\partial t} =
  \nabla\times \left(\left[\vv - \frac{3}{2} \Omega_0 x \yyy \right] \times \BB\right) - \frac{4\pi}{c} \nabla \times [\eta
  \JJ],
\label{induction} 
\end{equation}
where the current
\begin{equation}
\JJ \equiv \frac{c}{4 \pi} \nabla\times \BB,
\label{current}
\end{equation}
and the shearing box fluctuation velocity is $\vv$.  The $(3/2) \Omega_0 x \yyy$ term is the orbital shear.

So long as the resistivity remains constant in space,
\begin{equation}
\frac{\partial\BB}{\partial t} = \nabla\times \left(\left[\vv - \frac{3}{2} \Omega_0 x \yyy \right] \times \BB\right) -\eta \nabla \times [\nabla \times \BB].
\label{induction_2}
\end{equation}
The resistive term was integrated with the same scheme as the other operators.
The Pencil Code evolves an analogous equation for the magnetic vector potential ${\bm B} = \nabla \times {\bm A}$.

The thermal energy evolution includes a diffusive term based
on radiative transfer in the optically thick limit with Rosseland mean
opacities, as well as a thermal relaxation term to model the large
scale energy transfer away from the midplane of the disk.
The full thermal energy equation is thus
\begin{eqnarray}
\frac{\partial e}{\partial t} =& - \nabla \left(\left[\vv - \frac{3}{2} \Omega_0 x \yyy \right] e\right)- p_\mathrm{th}\nabla \cdot {\bm v}  \nonumber\\
  &+ \frac{4 \pi \eta}{c^2} J^2 -\nabla \cdot {\bm F} - C_v \frac{T-T_0}{\tau_0}
\end{eqnarray}
where $p_\mathrm{th}=(\gamma-1)e=\rho k_B T/\bar{m}$ is the thermal pressure, 
$C_v=(\gamma-1)^{-1} k_B/\bar{m}$ the heat capacity at constant volume,
$T$ is the gas temperature, $T_0$ the reference temperature, $\tau_0$ the thermal relaxation time, 
$k_B$ the Boltzmann constant, $\bar{m}$ the mean mass per particle,
 and 
with the Rosseland mean opacity thermal diffusion flux ${\bm F}$ given by
\begin{equation}
{\bm F} = -\frac{16 \sigma_B T^3}{3 \rho \kappa} \nab T\ ,
\label{radflux}
\end{equation}
where  $\sigma_B$ is the Stefan--Boltzmann constant, $\rho$ the density and $\kappa$ the Rosseland mean opacity.

The thermal diffusion treatment is motivated by the high optical depth
expected over length scales of the MRI unstable wavelength expected
for the outer edge of the MRI active region in the dusty inner disk.
In this region $\lambda_{\text{MRI}} \simeq H$, so the optical depth
of an MRI wavelength is comparable to that of the disk itself.  For
our parameters, $\lambda_{\text{MRI}}$ has an optical depth $\tau
>15,000$.  We choose the thermal relaxation timescale to be an orbital
period $\tau_0= 2 \pi/\Omega_0$ (Table~\ref{table_parameters}).  We
make this choice to probe the coolest reasonable disk state, which
corresponds to one that can cool so quickly that it approaches
marginal gravitational stability.  
The thermal relaxation rate is chosen to be  fast, so that it strongly limits 
the heating resulting from the energy input by the background shear.
Additionally, we can place an upper limit on the physically plausible cooling rate
by recognizing that the model is aimed to produce the environment at the inner edge of a dead zone.
If the cooling rate of the neighboring dead zone was much faster than $\tau_0=\Omega_0^{-1}$, 
the region would become gravitationally unstable \citep{2014MNRAS.438.1593R}, 
and would thus be prone to generating strong spiral waves, strongly altering the global disk structure.
As such, we can argue that this region of the disk must, to be consistent with an 
MRI-unstable quasi-steady state at the inner edge of the dead zone, 
have a cooling timescale longer than $\tau_0=\Omega_0^{-1}$.
Therefore, our choice of  $\tau_0=\Omega_0^{-1}$ should be a conservative one, and should result in a
 model which underestimates the temperature fluctuations.

\subsection{Initial Conditions}
\label{initial_conditions}

\begin{table}
\caption{Parameter Values}
\begin{center}
\begin{tabular}{lll}
\hline
& Parameter   & Value \\
\hline
$\rho_0$ & Initial density & $10^{-9}\ \mathrm{g\ cm^{-3}}$ \\
$T_0$ & Background temperature & $950\ \mathrm{K}$ \\
$L_x$  & Box size in $x$ & $0.3 \ \mathrm{AU}$ \\
        &                & $4.85H$ \\
$\Omega_0$ & Orbital frequency & $2\pi\ \mathrm{ yr^{-1}}$ \\
$r_0$ & Shearing box position & $1 \ \mathrm{AU}$\\
$\gamma$ & Gas adiabatic Index & $1.5$ \\
$\bar{m}$ & Gas mean particle mass & $2.33\ \mathrm{amu}$\\
$\eta$ & Ohmic resistivity $c^2/4\pi\sigma$ & $8.9\times10^{14}\ \mathrm{cm^2\ s^{-1}} $\\
           &                                                              & $5.2\times10^{-3} \Omega H^2$\\
$\beta_0$ & Initial plasma beta & $750$ \\
$v_{A0}$   & Initial Alfv\'en speed & $9.5\times10^3\ \mathrm{cm\ s^{-1}} $\\
                 &                                & $5.2\times10^{-2} \Omega H$\\
$\Lambda_0$ & Initial Elsasser number & 0.5 \\
$\kappa$ & Rosseland mean opacity & $20\ \mathrm{cm^2\ g^{-1}}$ \\
$\tau_0$ & Thermal relaxation time & $1\ \mathrm{yr}$ \\
$\lambda_{\rm MRI}$ & MRI fastest growing mode & $5.7\times10^{-2}\ \mathrm{AU}$ \\
                                   &                                            & $0.92H$\\
\end{tabular}
\end{center}
\label{table_parameters}
\end{table}

We chose the parameters listed in Table~\ref{table_parameters} to 
approximate a disk that is marginally MRI active
at 1~AU. Conceptually, these parameters can be thought of as 
corresponding to a region at the inner edge of a dead zone.  
Unusually for astrophysical magnetohydrodynamics, effects of the
microphysical resistivity are resolvable at the inner edge
of the dead zone (unlike the unresolvable microphysical viscosity) because there the MRI is marginally
super-critical and numerical simulations can clearly show the growth
of magnetic fields in an MRI dynamo.

The marginal instability criterion for the MRI is that the most unstable 
wavelength of the MRI be comparable to the disk
scale height.  
In our case, the MRI wavelength $\lambda_{\text{MRI}}$ and the effective scale height of the disk $H \equiv
c_s/\sqrt{\gamma} \Omega$ are within $10\%$ of each other. 
This means that even though we have performed unstratified numerical simulations, we can link
our length scales to those of stratified simulations near MRI-criticality.
With these parameters, the initial net vertical magnetic field corresponds to an initial plasma $\beta_0=750$.
This 
    is
low compared to many existing models with zero resistivity.
However, in this very resistive regime the wavelength of the fastest growing MRI mode increases with $\beta_0$.
Accordingly, a relatively strong initial field is needed
 to make the most unstable length scale match the nominal disk scale height.
The wavenumbers corresponding to the vertical heights of the shearing
box domains used in this study are noted in  
\Fig{fig_dispersion} along with the dispersion relation curve for the MRI.

While temperature is not strictly diffused, for the parameters in Table~\ref{table_parameters},
extrapolating from \Eq{radflux} and using $T=(\gamma-1)\bar{m} e/\rho k_B$, an effective thermal temperature  diffusion coefficient can be estimated as
\EQ
\mu \equiv \frac{16 (\gamma-1)  \bar{m} \sigma_B T_0^3}{3 \rho_0^2 \kappa k_B} \simeq 1.8 \times 10^{14}\  \frac{\text{cm}^2}{\text{s}},
 \label{therm_diff_eff}
 \EN
where $\sigma_B$ is the Stefan--Boltzmann constant,
and the other parameters are as defined in
Table~\ref{table_parameters}.  When measured with respect to the MRI wavelength and orbital timescale of our setup, we have
\EQ
\mu \sim \frac{\lambda_{\text{MRI}}^2 \Omega_0}{1000}.
\EN
Accordingly, the radiative cooling time for a structure with the size
of the most unstable MRI wavelength is approximately $130$ yr,
far longer than the imposed cooling time of $1$ yr.  This means that large scale thermal structures are dissipated
by the  thermal relaxation cooling term, but small scale
structures (less than $0.1 \lambda_{\text{MRI}}$) cool radiatively. 

\subsection{Runs}
We used the Pencil Code to perform two sequences of runs. 
The first  sequence uses a cubic volume with resolutions from $64^3$
to $512^3$, with
 $5 \lambda_{\text{MRI}}$ per box
length. Our highest two resolutions are 
 equivalent to $50$ and $100$ zones per scale height, respectively.
The higher resolution runs were
initialized from the next highest resolution run.
The $64^3$ simulation was run for $t=105$ orbits. At $t=45$ orbits, it was remeshed to a resolution of
 $128^3$ and this was also run to $t=105$ orbits.
 At $t=60$ orbits, the  $128^3$ run was remeshed to $256^3$ and this was run to $t=105$ orbits.
 Finally, at $t=75$ orbits, the  $256^3$ run was remeshed to $512^3$ and this was run to $t=105$ orbits.
The second sequence of runs repeats the same parameters with a slab volume one quarter the height, 
that is with aspect ratio $L_x:L_y:L_z$ of $1:1:(1/4)$.
Thus, these runs use a shearing box with a height of approximately one scale height.
These runs preserved the same cell-size as the first sequence of runs, so the highest resolution was $512^2 \times 128$.
Otherwise the procedures and analysis of this sequence of runs follows
    the cubical runs.

The comparison runs done with Athena 
used the cubical grid and
did not use remeshing. Both started from $t=0$ and ran for 50 orbits.
Resolutions of $128^3$ and $256^3$ are reported here.
The setup is otherwise identical to that used with the Pencil Code.

\section{Results}
\label{sec_results}

\subsection{Global Averages}
\begin{figure*}[t!]\begin{center}
\includegraphics[width=\textwidth]{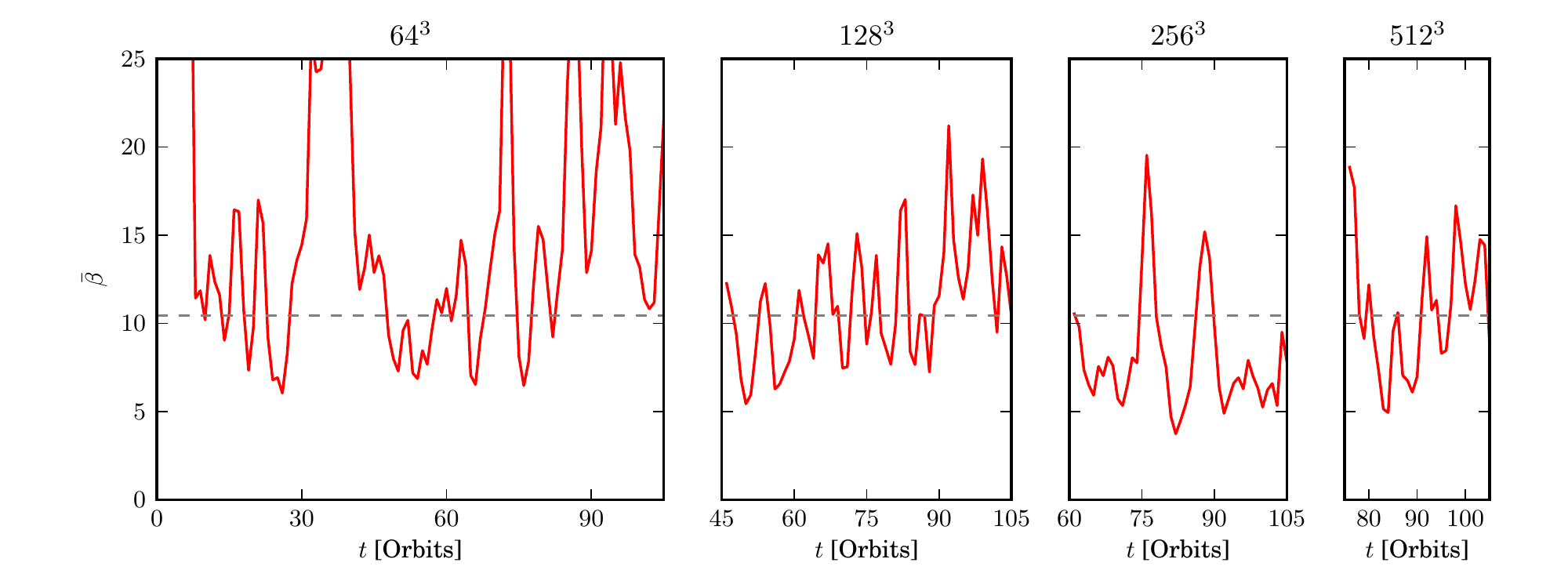}
\includegraphics[width=\textwidth]{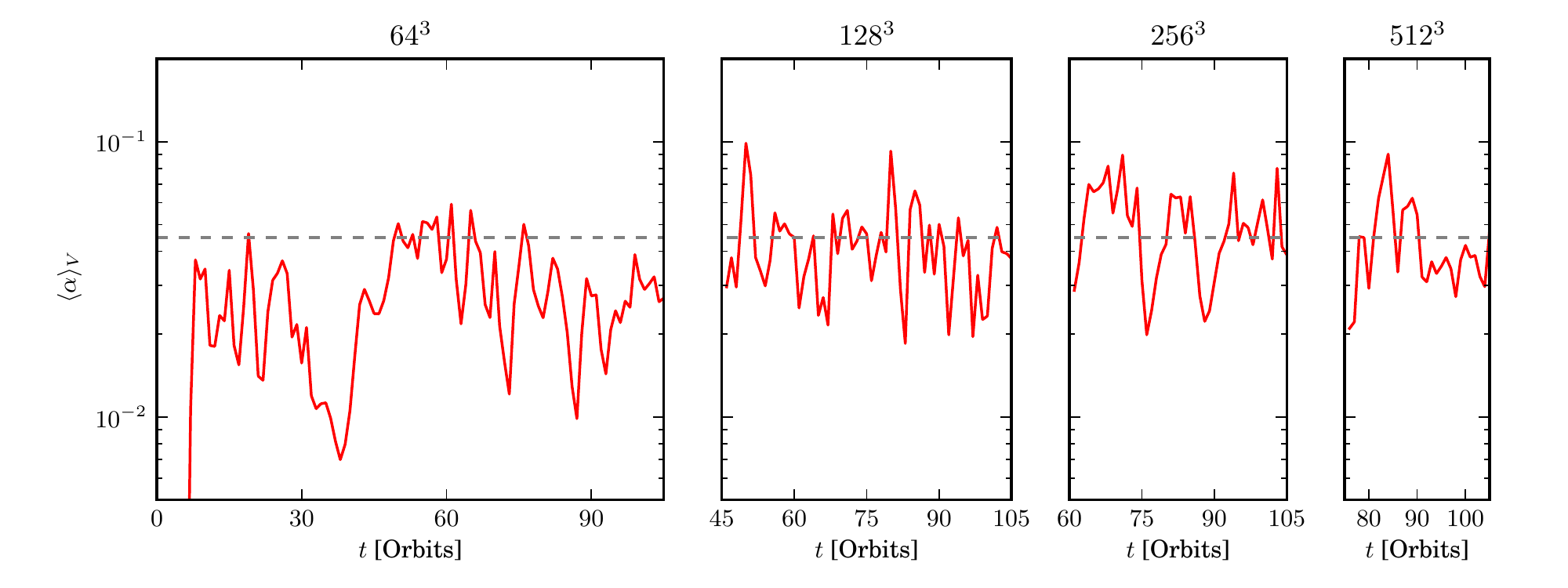}
\end{center}\caption{ Volume averages of the plasma $\overline{\beta}$ and total stress 
$\langle \alpha \rangle_V$ for the cubical Pencil Code runs.  Note that the scale of the time-axis
is the same for all panels, but the higher resolution runs have
shorter total run-time, as they were remeshed from the next highest resolution run.
The average value for the
highest resolution runs are shown by the dashed lines to guide the
eye.
\label{beta_alpha_volavg} }
\end{figure*}

\begin{figure*}[t!]\begin{center}
\includegraphics[width=\textwidth]{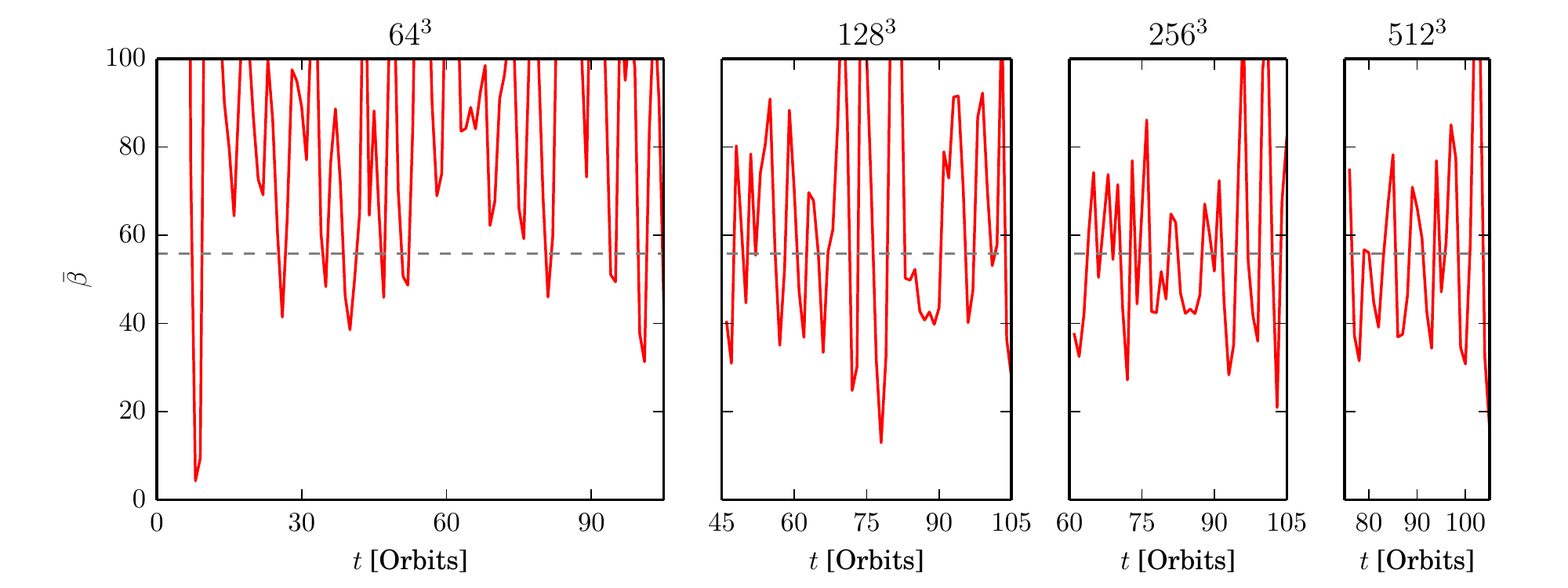}
\includegraphics[width=\textwidth]{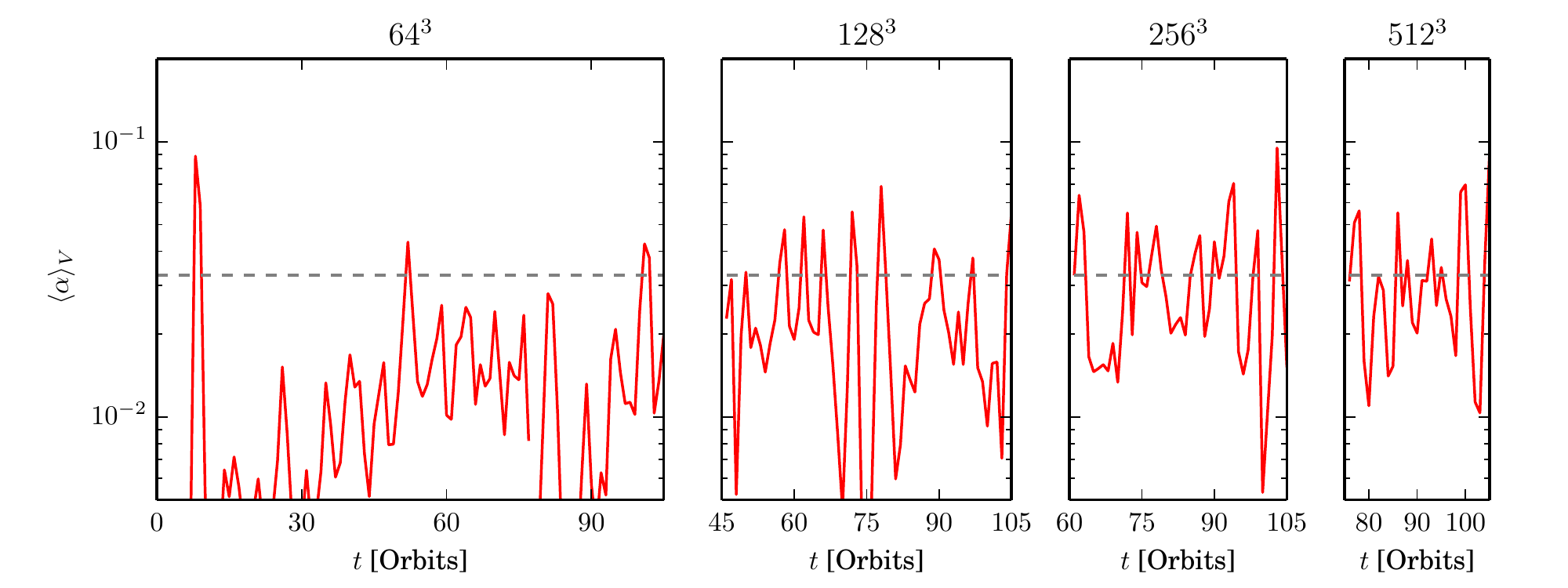}
\end{center}\caption{ 
Volume averages of the plasma $\overline{\beta}$ and total stress 
$\langle \alpha \rangle_V$ for the slab Pencil Code runs.  Note that the scale of the time-axis
is the same for all panels, but the higher resolution runs have
shorter total run-time, as they were remeshed from the next highest resolution run.
The average value for the
highest resolution runs are shown by the dashed lines to guide the
eye.
\label{beta_alpha_volavg_run8} }
\end{figure*}

\begin{figure*}[t!]\begin{center}
\includegraphics[width=\textwidth]{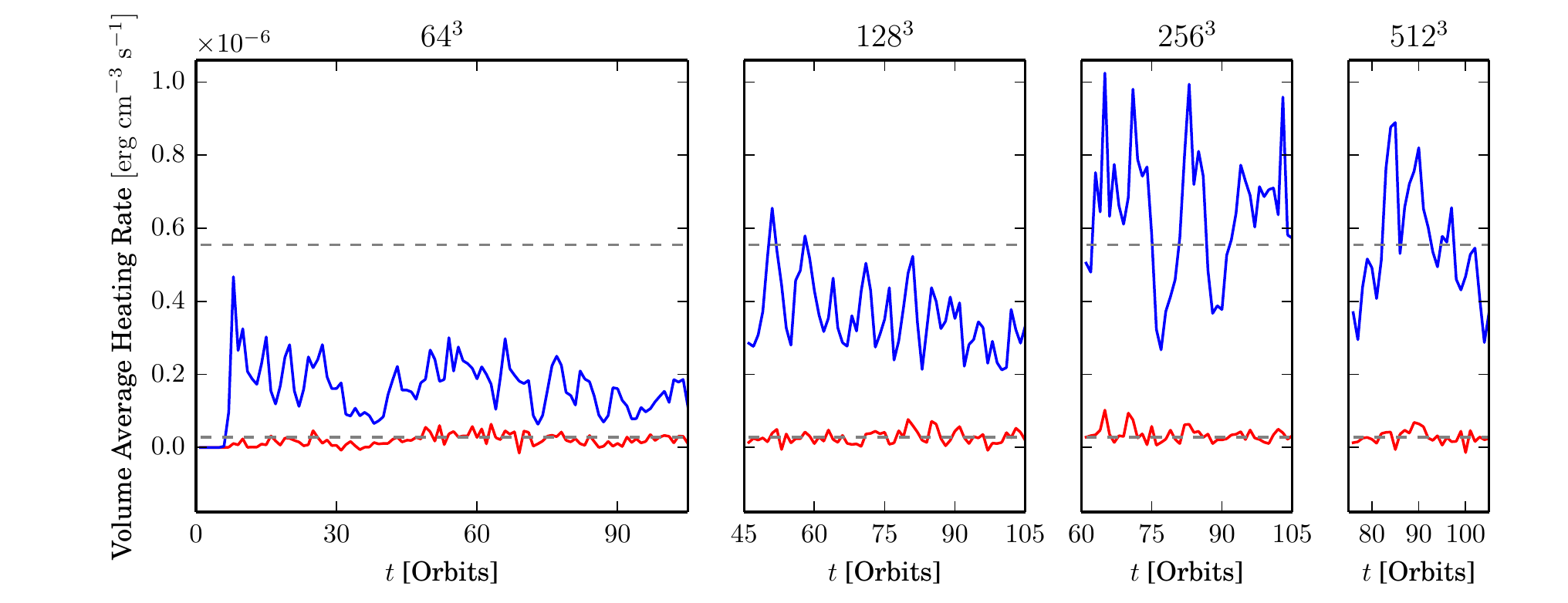}
\includegraphics[width=\textwidth]{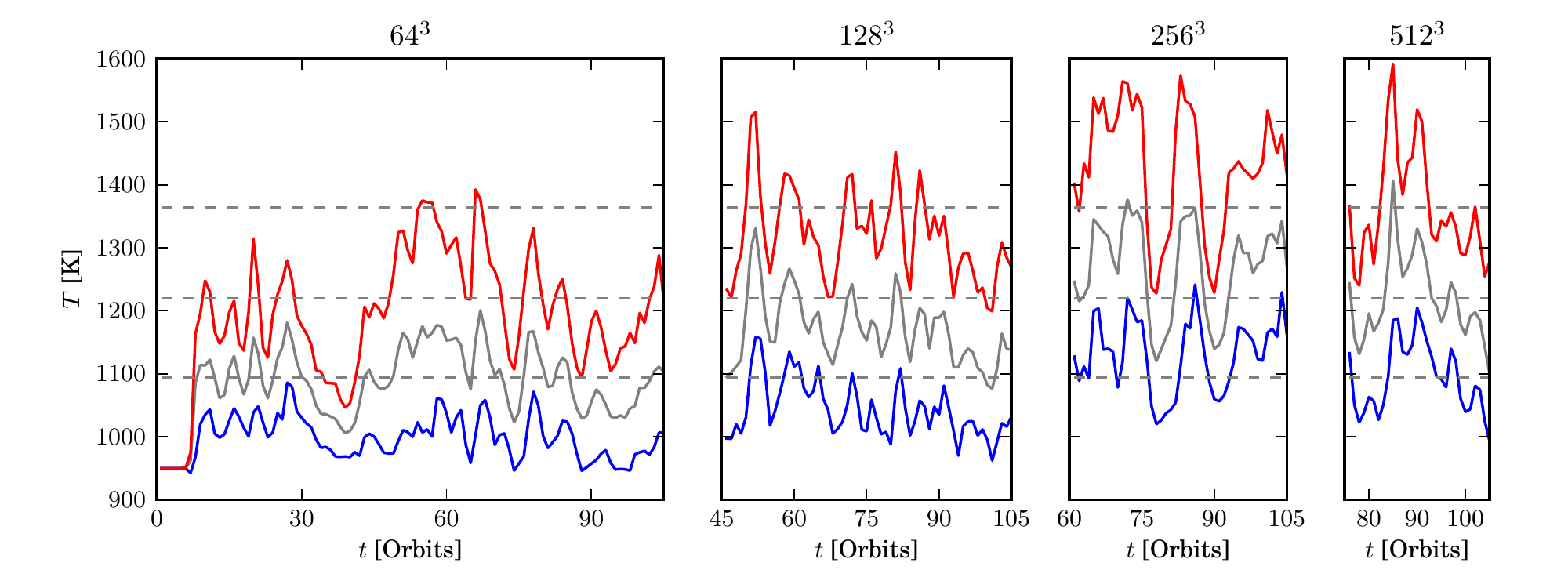}
\end{center}\caption{ Upper panel: volume averaged heating rate for
  the cubical Pencil code runs.  
Resistive heating in blue
and compressive heating or cooling in red. 
Lower panel: ninetieth percentile (red), fiftieth percentile (gray)
and tenth percentile (blue) temperatures for the same runs.  Higher resolution runs have shorter run times as they were remeshed from the next highest resolution run.
For each data series, the average value for the highest resolution runs are shown by the dashed lines to guide the eye.
\label{res_check} }
\end{figure*}

\begin{figure*}[t!]\begin{center}
\includegraphics[width=\textwidth]{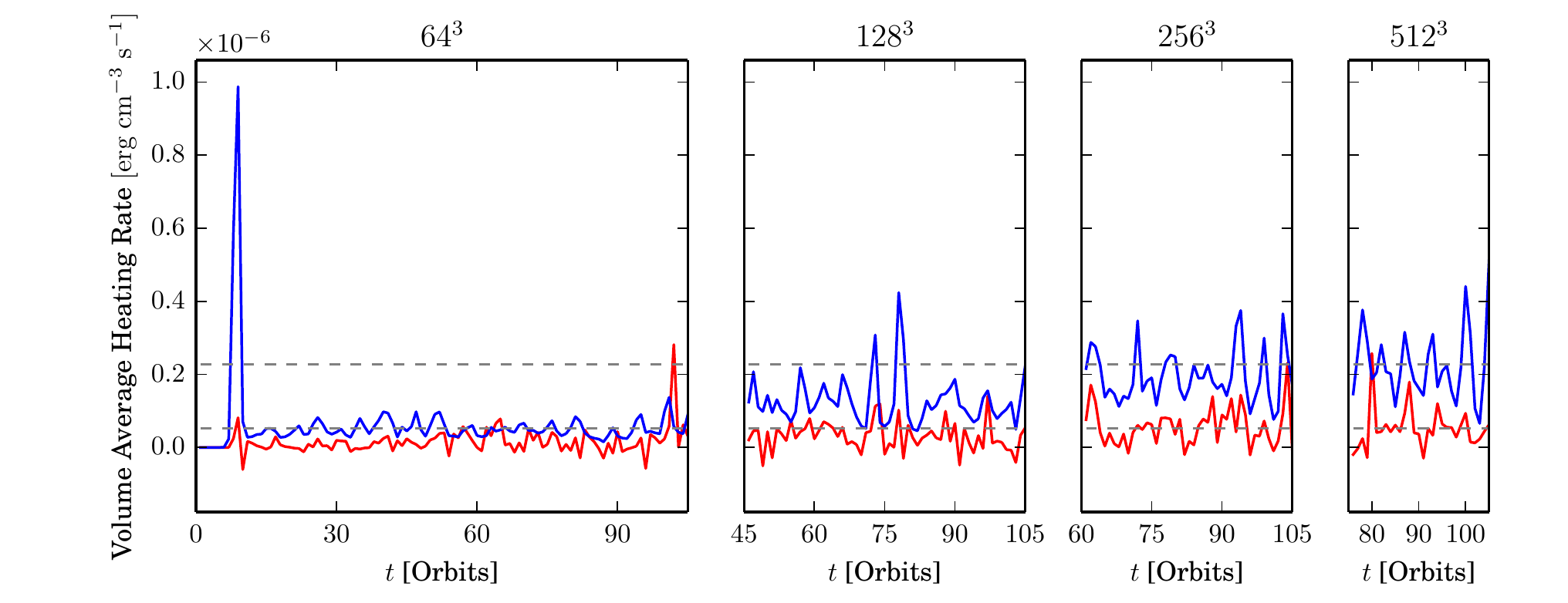}
\includegraphics[width=\textwidth]{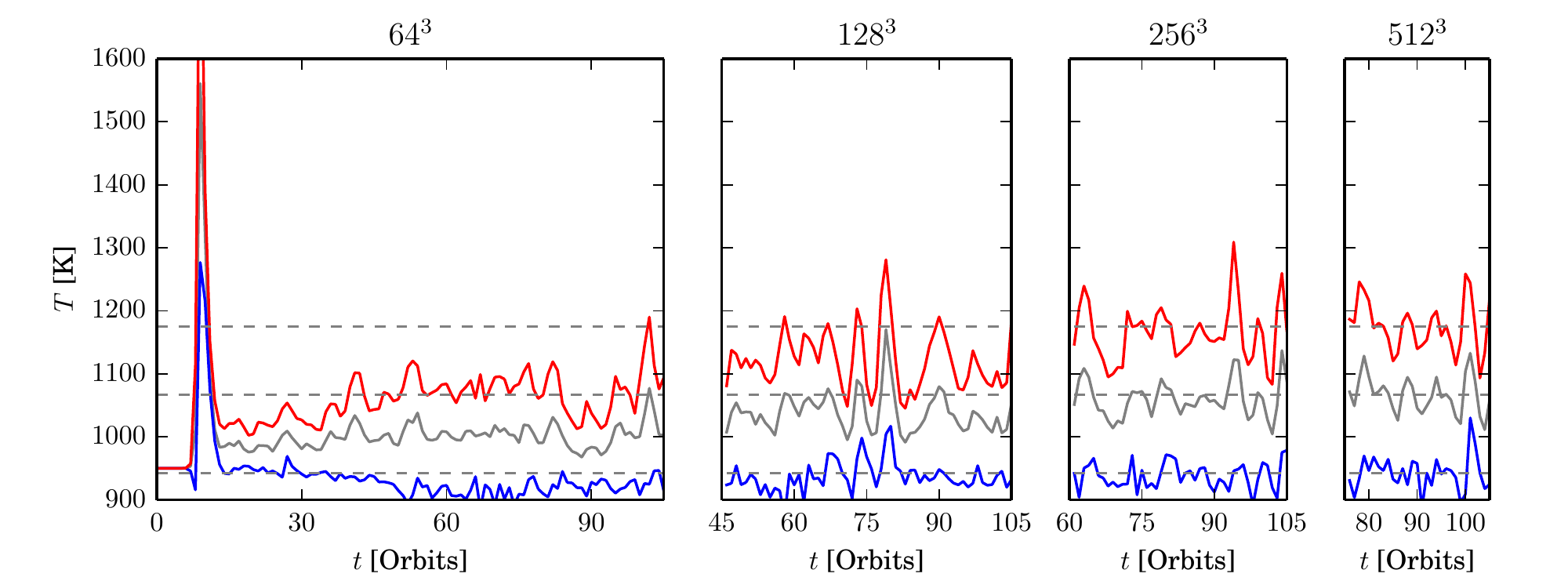}
\end{center}\caption{ 
Upper panel: volume averaged heating rate for the slab Pencil code runs.  
Resistive heating in blue
and compressive heating or cooling in red. 
Lower panel: ninetieth percentile (red), fiftieth percentile (gray)
and tenth percentile (blue) temperatures for the same runs.  Higher resolution runs have shorter run times as they were remeshed from the next highest resolution run.
For each data series, the average value for the highest resolution runs are shown by the dashed lines to guide the eye.
\label{res_check_run8} 
}
\end{figure*}

We first consider the behavior of the global average quantities in our models.
In the top panels of Figures~\ref{beta_alpha_volavg}
and~\ref{beta_alpha_volavg_run8} we show time series of the volume
average plasma beta 
\EQ
\overline{\beta}=\frac{8 \pi \langle p_\mathrm{th} \rangle_V}{\langle B^2 \rangle_V},
\EN
where $\langle \cdots \rangle_V$ denotes volume averaging, for our
different resolution runs.  
The volume average of the magnetic field was taken separately to avoid
having this average skewed by very high values of $\beta$ within
demagnetized reconnecting regions.
Our cubical models settle at
values of $\overline{\beta} \sim 10$, while the slab models only reach
$\overline{\beta} \sim 50$.
Similarly, the bottom panel of both figures shows the volume averaged 
\cite{1973A&A....24..337S} $\alpha$ turbulent viscosity
parameter: 
\EQ
\alpha =\frac{1}{p_\mathrm{th}} \left(\rho v_x v_y - \frac{B_x B_y}{4\pi} \right).
\EN
The volume averaged value of this parameter
settles at a value of about $\langle \alpha \rangle_V=4 \times 10^{-2}$
in the cubical case and 
 $\langle \alpha \rangle_V=3 \times 10^{-2}$
in the slab case.

In Figures~\ref{res_check} and~\ref{res_check_run8} 
 we show temperature and heating data from the
Pencil Code runs.  Two facts jump out. First, there is an order
unity difference between the tenth and ninetieth percentile temperatures at
all times.  Second, the dominant energy
dissipation mechanism is resistive rather than compressive or shock
dissipation, because when averaged over the volume, the 
adiabatic heating and cooling approximately
cancel. 
This occurs even though the compressive heating has much stronger peak values
than the resistive heating, as we show in the next section.

\subsection{Two-dimensional Slices}

\begin{figure*}[t!]\begin{center}
\includegraphics[width=0.8\linewidth]{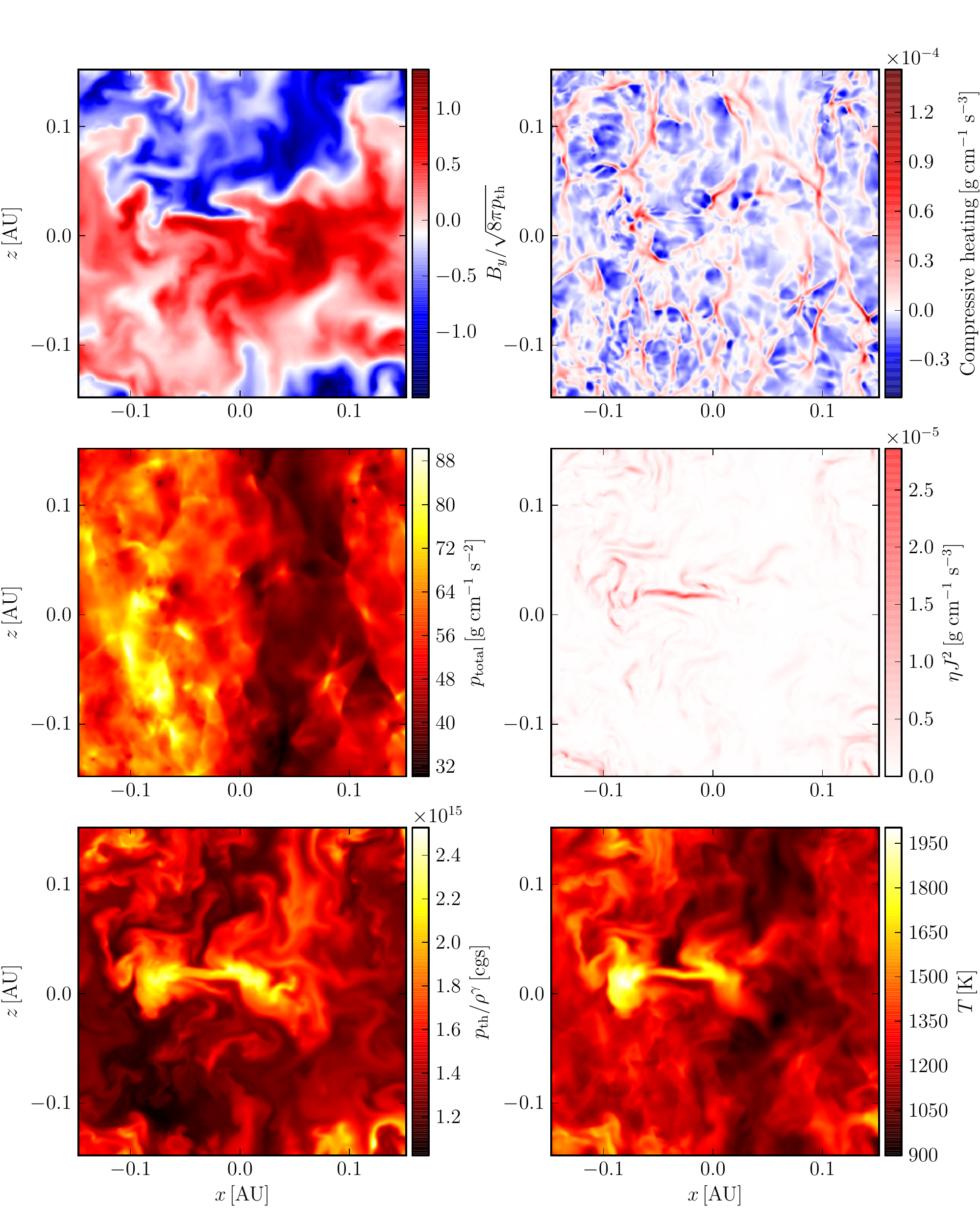}
\end{center}\caption{ Data from a radial--vertical slice of the cubical, $512^3$
  Pencil Code run at time $t=83$ orbits and azimuth $y=0\ \mathrm{AU}$, cutting through the
  largest current sheet on the grid.
Top left panel: azimuthal magnetic field, showing a large scale vertical dependence.
Middle left panel: total pressure, showing a large scale horizontal dependence known as a zonal flow.
Bottom left panel: entropy, showing a maximum in the current sheet
centered near $(-0.03~\text{AU}, 0.015~\text{AU})$, so the temperature 
of the current sheet is not primarily due to adiabatic compression.
Top right panel: compressive heating and cooling, showing that shocks
permeate the domain, but the current sheet does not stand out. 
Middle right panel: Resistive heating, clearly showing the current
sheet, using the same color scale as the top right panel.
Bottom right panel: temperature, showing that the temperature near the
current sheet far exceeds the background.
\label{Slices_full} }
\end{figure*}

To more closely investigate the source of the temperature variations, we
turn to a study of the morphology of current sheets.  We find that the
heating is usually dominated by one or two major sheets at any given
time.  In \Fig{Slices_full} we show a set of slices at constant
azimuth through our highest resolution cubical Pencil Code run.  In the slice
chosen, there is a current sheet lying in the radial-azimuthal plane,
perpendicular to the slice plane, and
visible in all variables except the total pressure, although only
barely visible in the adiabatic heating and cooling panel.  This
figure makes clear that, even though the volume average $\overline{\beta} \simeq 8$ at
the orbit in question, the local minimum ratio associated with the peak magnetic field
is $\beta_p \simeq 1$.

Furthermore, the largest temperature variation generally
traces the highest resistive dissipation in the current sheet
structure, although there is enough difference between the two to make
clear that it is time-averaged, rather than instantaneous, heating
that determines the temperature perturbation.
Although the compressive heating is high along the weak shocks filling the domain,
 these are accompanied by large regions of expansion cooling.
Hence, as was seen in the volume average shown in the upper panel of Figure~\ref{res_check},
 the heating is dominated by the resistive dissipation.

\begin{figure*}[t!]\begin{center}
\includegraphics[width=0.95\linewidth]{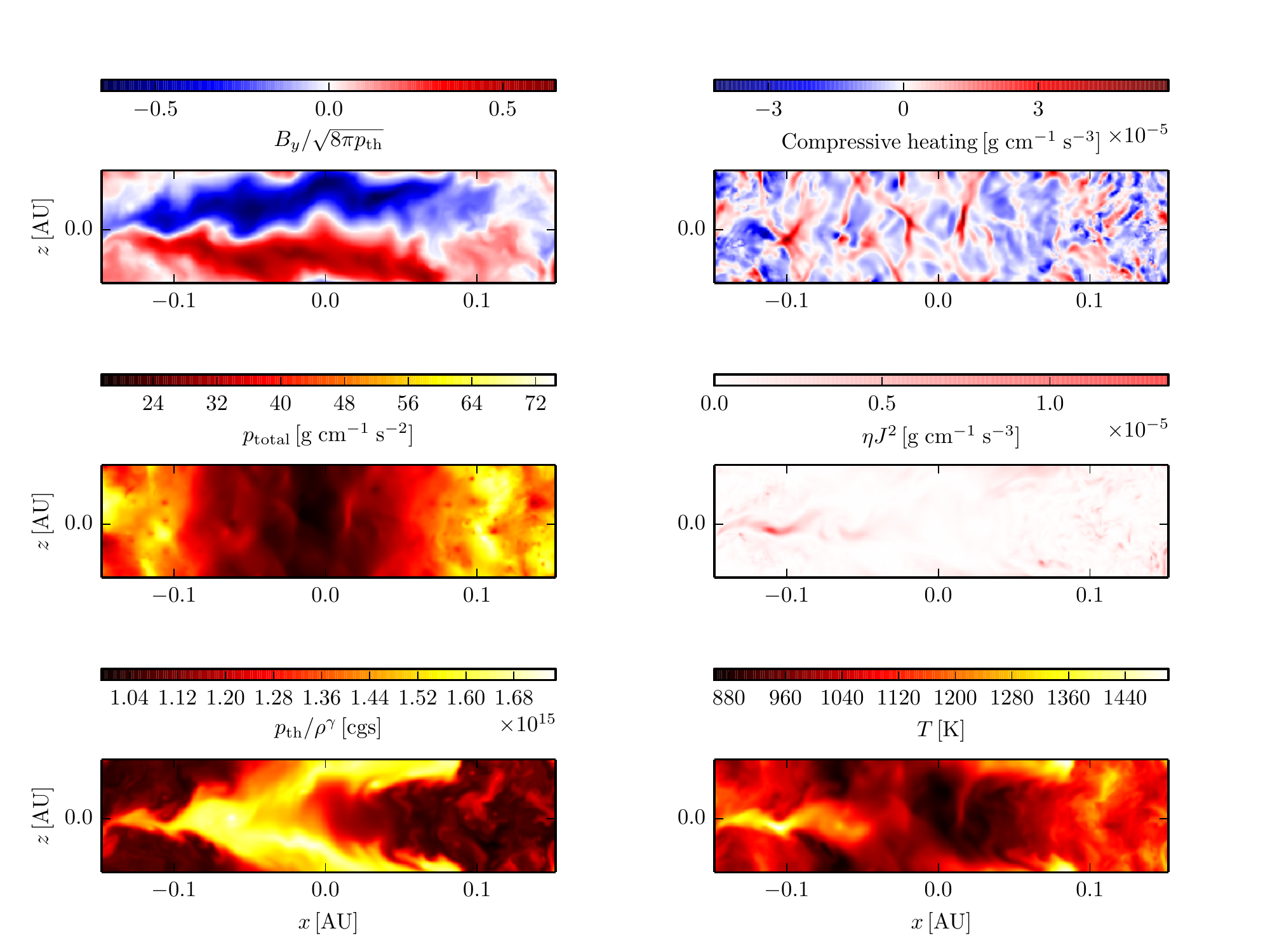}
\end{center}
\caption{ Data from a radial-vertical slice of the $512^2
  \times 128$ slab
  Pencil Code run at time $t=105$ orbits and azimuth $y=0.15\ \mathrm{AU}$, 
  cutting through the
  largest current sheet on the grid.
Top left panel: Azimuthal magnetic field, showing a large scale vertical dependence.
Middle left panel: Total pressure, showing a large scale horizontal dependence known as a zonal flow.
Bottom left panel: Entropy, showing a maximum in the current sheet
centered near $(-0.1~\text{AU}, 0.0~\text{AU})$, so the temperature 
of the current sheet is not primarily due to adiabatic compression.
Top right panel: Compressive heating and cooling, showing that shocks
permeate the domain, but the current sheet does not stand out. 
Middle right panel: Resistive heating, clearly showing the current
sheet, using the same color scale as the top right panel.
Bottom right panel: Temperature, showing that the temperature near the
current sheet far exceeds the background.
\label{Slices_full_run8} }
\end{figure*}

 Similarly, \Fig{Slices_full_run8} shows a set of slices at constant
azimuth through our highest resolution slab run.
The qualitative pattern of dominant azimuthal field bundles is similar, and a strong, 
hot current sheet structure can be seen.
Consistent with the results for the globally averaged quantities, the  values are less extreme in this 
vertically restricted domain than in the cubic one: the volume averaged $\overline{\beta} \simeq 20$
and the local minimum $\beta$ associated with the peak magnetic field is $\beta_p \simeq 1$.

\begin{figure}[t!]\begin{center}
\includegraphics[width=\columnwidth]{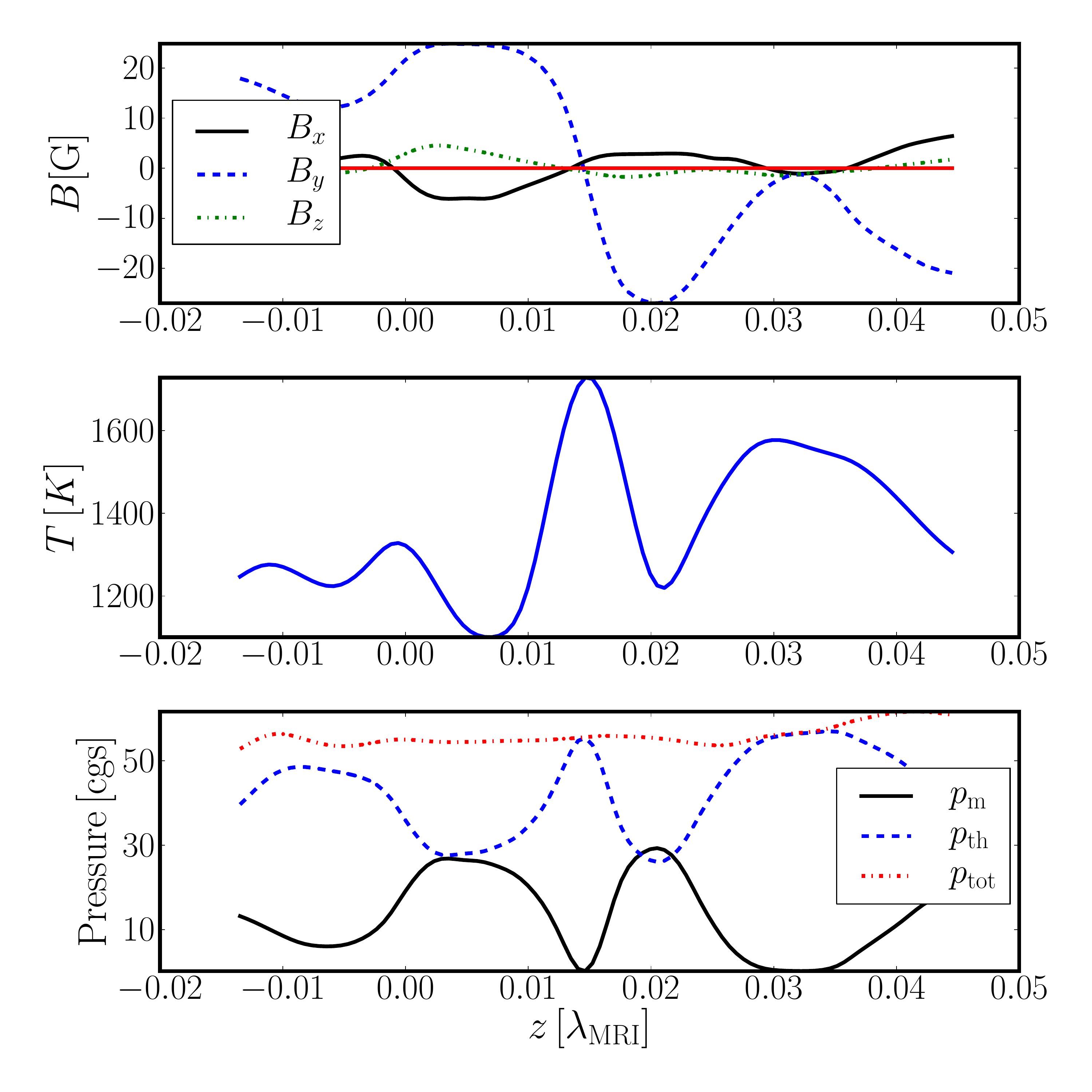}
\end{center}\caption{ 
Data from a radial--vertical slice of the current
  sheet at $(0.03~\text{AU},
0.015~\text{AU})$ in the highest resolution Pencil Code run at the
same time as Figure~\ref{Slices_full}.
Vertical line cut through the current sheet.  Top panel: magnetic
field ($\xx$ in black, solid; $\yy$ in blue, dashed;
$\zz$ in green, dash-dotted). Note the anti-correlation of $B_x$ and $B_y$ as the latter is the product
of shearing the former.  Middle panel: temperature. Bottom
panel: magnetic pressure (black, solid), thermal pressure (blue,
dashed), and total pressure(red, dash-dotted). This shows the degree
of pressure balance across the current sheet.
\label{line_cut} }
\end{figure}

This difference between the local $\beta_p$ and volume averaged
$\overline{\beta}$ is straightforward.
The strongest magnetic field structures of the MRI are generated by orbital shear, which
stretches radial magnetic field lines into nearly azimuthally constant bundles of strong,
azimuthally directed magnetic flux.
If the magnetic
field is dominantly contained in azimuthally constant structures varying along a single large scale
wave-vector $\kk$, we would expect $\beta_p \simeq \overline{\beta}/4$ because we
need to average over $\sin(k_x x)^2 \sin(k_z z)^2$.  If $\beta_p$ is
of order unity, its value is expected to further drop because the
magnetic field bundles will have low thermal pressure compared to the
volume averaged thermal pressure (because the magnetic pressure pushes
fluid out of the high magnetic field regions).  In \Fig{line_cut} we
show a vertical cut through the current sheet identified in \Fig{Slices_full} that demonstrates
that this inverse magnetic and thermal pressure correlation
 occurs in large current sheets found
in the highest resolution aspect ratio cubical model.

These figures also show
that even though the simulation has $\beta_p \sim 1$ and hence drives
mildly supersonic flows, the large temperature variations are due to
resistive heating.  Regions heated by the hydrodynamical shocks quickly reexpand
and adiabatically cool, leaving little imprint on the temperature structure
of the gas.
\label{currentsheets}

\section{Model Reliability}
\label{sec_Resolution}
Our conclusions depend on resolving the heating within thin current
sheets.  We therefore have made a detailed study of the convergence
properties of our models, and confirmed our Pencil Code results by
comparison to models run with an independent numerical method.

\subsection{Global Averages}
 Figures~\ref{beta_alpha_volavg} and \ref{beta_alpha_volavg_run8} compare the time
variation of $\overline{\beta}$,  $\langle \alpha \rangle_{V}$,
in runs of varying resolution for the two aspect ratios.  The average value for the
highest resolution runs are shown by the dashed lines to guide the
eye. 
Similarly Figures~\ref{res_check} and~\ref{res_check_run8}
compare the temperature and heating rates.
Episodes of very low
field strength are common for the lower resolution runs, which are
less well resolved, and hence subject to stronger numerical
resistivity.  Our simulations
appear to have converged to at least the intrinsic time variation in
these averaged quantities for resolutions of
$128^3$ or larger.  However, the temperature and the resistive heating
rate converge only for the two highest resolutions
($256^3$ and $512^3$, or $50$ and $100$ zones 
per $\lambda_{\text{MRI}}$ or equivalently per scale height.). 

The motivation for the thinner domain is to avoid the cycles of strong channel 
flow and intermittency that have been observed to be exacerbated by the use of a cubic domain \citep{2008A&A...487....1B}.
Indeed, the variation on $\sim 10$ orbit timescale appears less in
these runs than in the cubical ones.
However, channel mode-like pairs of azimuthal field bundles still
dominate the magnetic field, as can be seen in the top left panel of
Figure~\ref{Slices_full_run8}.
The heating is less extreme for the thinner box, but the largest MRI
active scales are also truncated by the vertical height (see Figure~\ref{fig_dispersion}). 
Accordingly, the volume average $\bar{\beta}$ is on the order of five times higher in these runs.

We explain the increase in volume averaged heating rate with
resolution by reference to the theory described in
Section~\ref{OrbShear}.  Magnetic fields are current sources: $\JJ \propto
\nab\times \BB$. The MRI often produces oppositely directed azimuthal magnetic
field bundles. Given two such bundles
with strength $B_0/2$ separated by a distance $L$, the current flowing
between them scales as $J \sim B_0/L$.  However, resistive dissipation
scales with the square of the current, $\eta |\JJ|^2$, so the net
resistive heating scales as $\eta B_0^2/L^2$.  If the current sheet is
not resolved the heating rate will be understated because the actual
value of $L$, set by the resolution, is larger than the physical one
set by $\eta$.

The high resolution required to resolve the current sheets, even at
the high resistivity edge of the dead zone,
can be understood from dimensional considerations.  Numerical resistivity
$\eta_n \propto u \times \Delta x$, where $u$ is the turbulent velocity. If the imposed
resistivity
\EQ
\eta \gg \eta_n, \label{res_requirement}
\EN
 then the physical resistivity $\eta$ dominates over
the numerical resistivity $\eta_n$ in \Eq{induction} and resistive effects are resolved.

Assuming equipartition between the turbulent kinetic and turbulent
magnetic energies, $u^2 \sim v_A^2$, so $\eta_n \sim v_A \times \Delta x$,
where $v_{A}$ is the Alfv\'en speed 
of the saturated magnetic field.  The ratio of $\eta_n$ for the saturated state to
$\eta_{n0}$ of the initial state is $\eta_n/\eta_{n0} \simeq  v_{A}/v_{A0} = (\beta_i/\beta_s)^{1/2}$ 
where $\beta_i$ is the plasma beta of the initial field, and $\beta_s$ is that
of the saturated field.  This ratio is 
often an order of magnitude or larger.

This can cause issues even near the relatively high resistivity edge of the dead zone where \Eq{res_requirement} is
best satisfied.  The MRI criticality condition is that the initial Elsasser number $\Lambda_0 =v_{A0}^2/\eta \Omega$ be of order unity, where $v_{A0}$ is the Alfven speed of the seed field.  The magnetic Reynolds number of the saturated state is
\EQ
 \Rm \equiv u^2 t_t/\eta,
\EN 
where $t_t$ is the turbulent turnover time, $t_t \simeq\Omega^{-1}$, 
and taking $u$ to be the maximum shearing-box fluctuation velocity.
In our case, the initial Elsasser number is  $\Lambda_0=0.5$ and 
at late times the turbulent magnetic Reynolds number is 
$\Rm \sim 100$,
 which is a non-trivial value of $\Rm$ to resolve numerically. 
Note that $\Rm/\Lambda_0 = (\eta_n/\eta_{n0})^2$, so the grid resolution needed to resolve the turbulent flow in the
saturated state is a factor of over 14 times more stringent than at
the initial time. All our runs, including the lowest resolution,
$64^3$ ones, resolved the initial state according to
\Eq{res_requirement}.  During the highest energy episodes, however,
even the highest resolution $512^3$ simulations did not quite satisfy
\Eq{res_requirement}, though only by a factor of about two.  To
understand the actual convergence behavior, we need to move beyond
that criterion's simple dimensional analysis.

\subsection{Decomposition of $|\JJ|^2$}

\begin{figure}
\begin{center}
\includegraphics[width=\columnwidth]{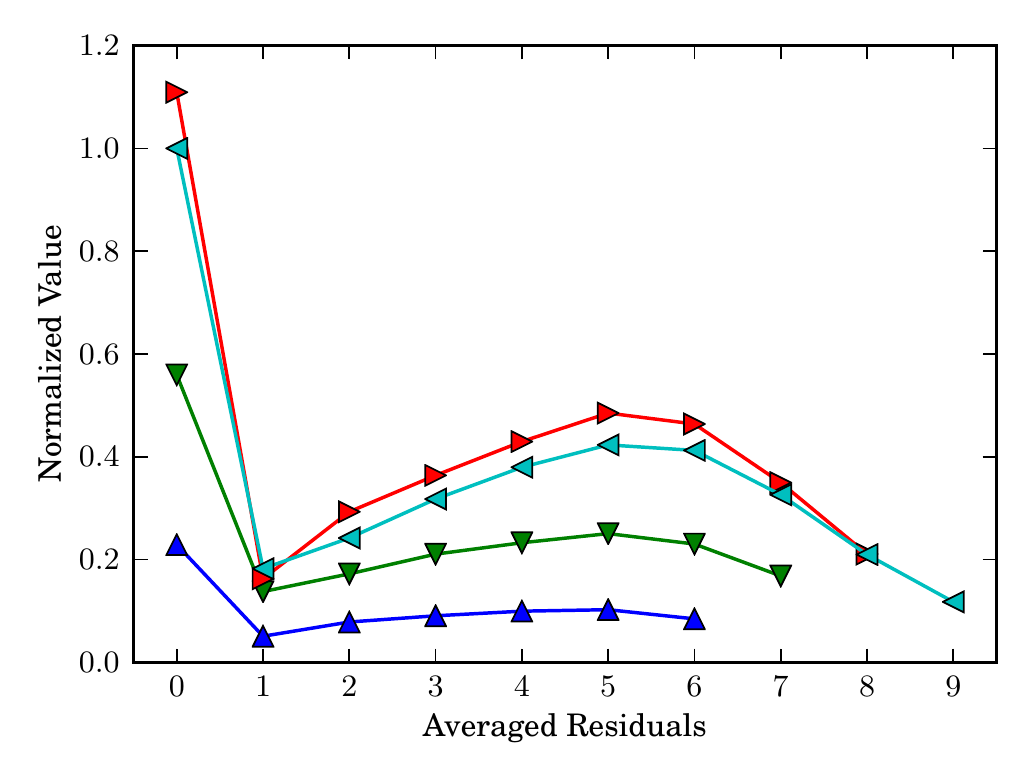}
\caption{
Time-averaged multiresolution decomposition of  $|\JJ|^2$ in cubical Pencil Code runs.
On the $x$-axis, $0$ is the $v^0$ residual, and $1$ through $9$ are
the $e^1$ to $e^9$ residuals corresponding to refined grid levels 1--9.
All values are averages of absolute values over the mesh and over
orbits 85--105, and are normalized to the $v^0$ coefficient of the $512^3$ resolution run.
The resolved peak in the residuals is at level $5$. Base grid
resolutions shown are $64^3$ ({triangle up, blue}),
$128^3$ ({triangle down, green}),
 $256^3$ ({triangle right, red}), and
 $512^3$ ({triangle left, cyan}).
}
\label{mr_j2} 
\end{center}
\end{figure}

\begin{figure}
\begin{center}
\includegraphics[width=\columnwidth]{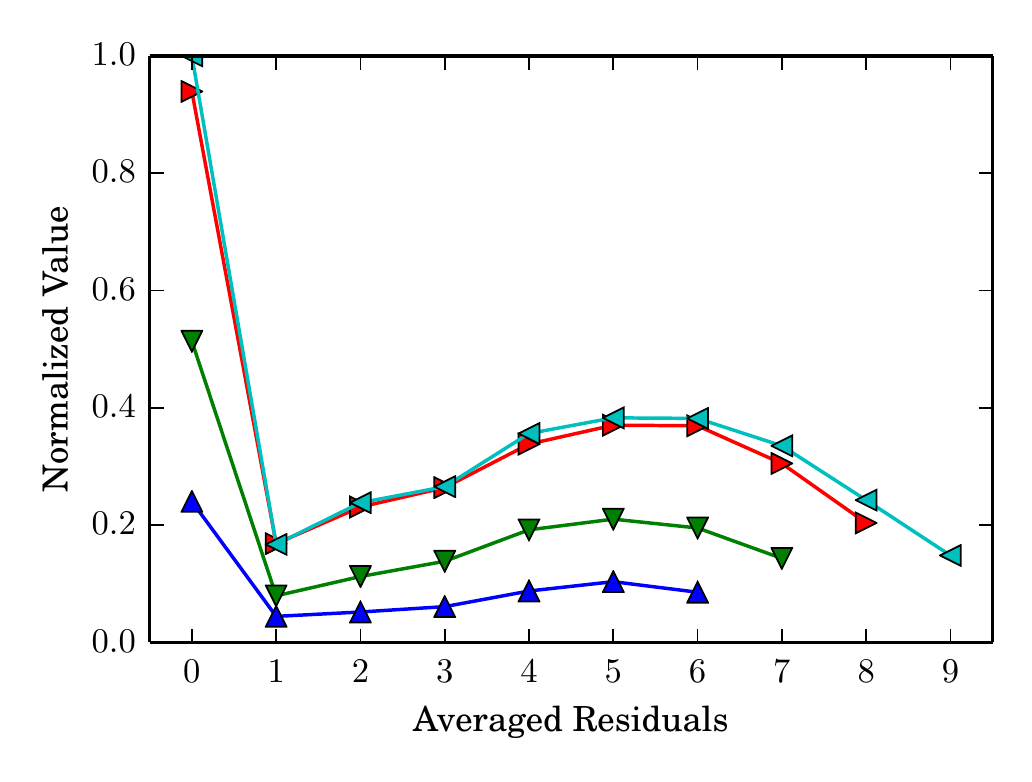}
\caption{
Time-averaged multiresolution decomposition of  $|\JJ|^2$ in slab Pencil Code runs.
The symbols and labels are the same as Figure~\ref{mr_j2}.
}
\label{mr_j2_run8} 
\end{center}
\end{figure}

To
better probe the resolution required to resolve resistive
dissipation we examine the structure of the current density.  
Our current sheets show multiple lengthscales: the full box length in azimuth, $\lambda_{\text{MRI}}$
in width and a far shorter length perpendicular to the sheet.  This rules out a simple Fourier analysis
of the current sheet structures.
Instead, to examine the convergence of the complex current sheet structures we decomposed $|\JJ|^2$ 
with a multiresolution decomposition \citep{harten94}.  

We successively coarse-grained the data, and subtracted the coarse-grained
data from the original.
We call the results of the subtraction residuals, although they
are described as error coefficients in other contexts. 
They measure how much structure there is on a given scale and hence at
what scale the dominant features of $|\JJ|^2$ exist.
The decomposition is conservative, in that the sum 
of the residuals 
on every level over the entire volume is zero.
To detect how much structure in the field resides on each level, 
 we examine the average absolute value of the residuals on each grid level.
  
 The transform used to calculate the residuals is formally
\EQ
v^L \longleftrightarrow \left\{ e^L,...,e^1, v^0 \right\}.
\EN
The function $v$ is on the finest mesh $v^L$, the
    residuals on each level $N$ are $e^N$,
and $v^0$ is the representation of the function $v$ on the coarsest mesh: a single point.
The transform uses a decimation operator to move the representation of the function $v$ from mesh level $N$ to 
a mesh with half as many points in each direction on level $N-1$ by 
\EQ
v^{N-1}_{i,j,k} = \frac{1}{8} \sum_{l,m,n=0}^1 v^N_{i+l,j+m,k+n} 
\EN
which is a simple average over a cube of eight neighboring points. 
In this way the volume integral of $v$ is conserved in the successively coarser representations.
Thus the most coarse representation $v^0$ is a scalar value which is equal to the volume average of $v^L$.
The residuals on each level are defined by the difference between the representation 
of the function $v$ on level $N$ and level $N-1$ as
\begin{eqnarray}
e^N_{2i+l,2j+m,2k+n} &= &v^N_{2i+l,2j+m,2k+n} - v^{N-1}_{i,j,k} \nonumber \\
&& l,n,m \in \{0,1\}\ .
\end{eqnarray}
The mean absolute value of the residual $e^N$ on level $N$, denoted
as $\langle |e^N| \rangle$ is then a measure of the total volume-integrated changes between
$v^{N-1}$ and $v^N$.

If we examine the set of values $\left\{ \langle |e^1| \rangle, ..., \langle |e^L| \rangle \right\}$, 
the location of the largest residual
corresponds to the grid scale where the original function $v^L$ changes most.
The single $|v^0|$ value
measures the volume averaged $|\JJ|^2$.
To account for the temporal variations of the system, we compute this decomposition once per orbit,
and report averages of the absolute values of the coefficients, normalized to the time average of
$|v^0|$ from the highest resolution simulation.

The results of this analysis are shown in Figures~\ref{mr_j2} and~\ref{mr_j2_run8}. 
The time
averages were performed over the interval $t=80$ to $t=105$.
The $|v^0|$ values show that the volume integrated $|\JJ|^2$ only approaches
 convergence at resolution of $256^3$ to $512^3$.
Both the cubical and slab simulations
show the same convergence behavior.
 The averaged residuals $e^N$ consistently peak at the scale
 corresponding to level 5;
 as resolution increases, the finest scales show decreasing $e^N$
 coefficients.  The strength of the residuals converges well at the
 highest resolutions.  Note that $e^5$ corresponds to the structure of
 $\J^2$ on a mesh with resolution $32^3$, i.e.~to a scale of
 approximately $\lambda_{\text{MRI}}/6$.  Hence, even though $|\JJ|^2$
 shows structure on all scales, our highest resolution simulations are
 well able to resolve the dominant scales of current sheet
 dissipation.

\subsection{Athena Comparison}

\begin{figure}[t!]\begin{center}
\includegraphics[width=\columnwidth]{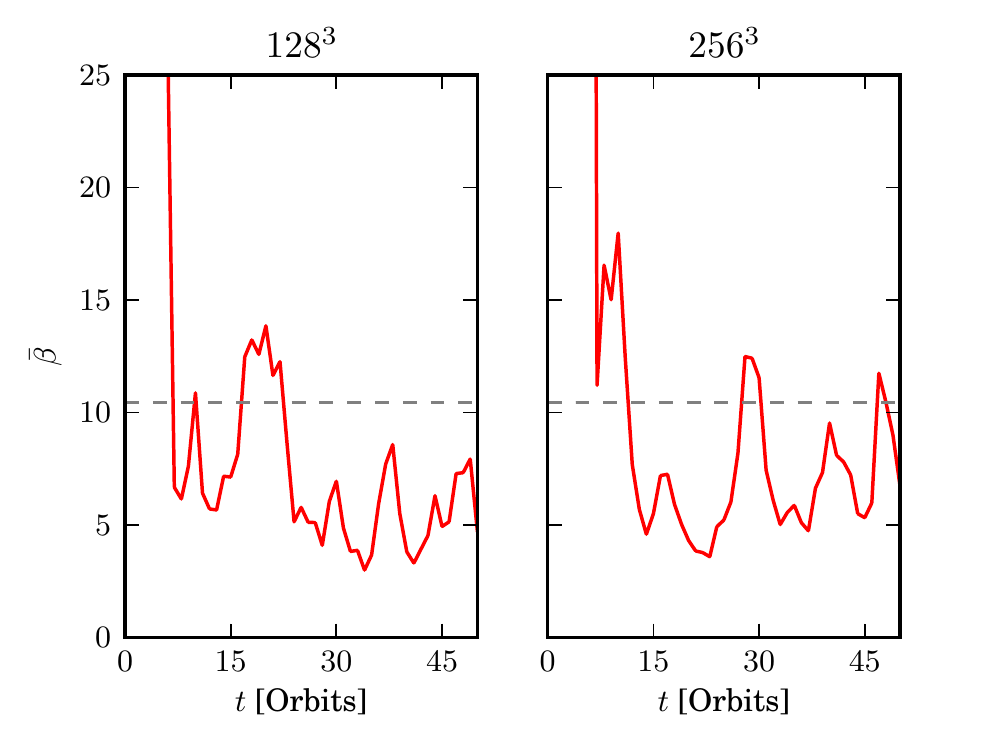}
\end{center}\caption{
Time series of $\ov{\beta}$ for the Athena runs.  Compare to \Fig{beta_alpha_volavg}, top panels, the dashed gray reference line is the same as shown there.
}
\label{abeta_volavg} 
\end{figure}

\begin{figure}[t!]\begin{center}
\includegraphics[width=\columnwidth]{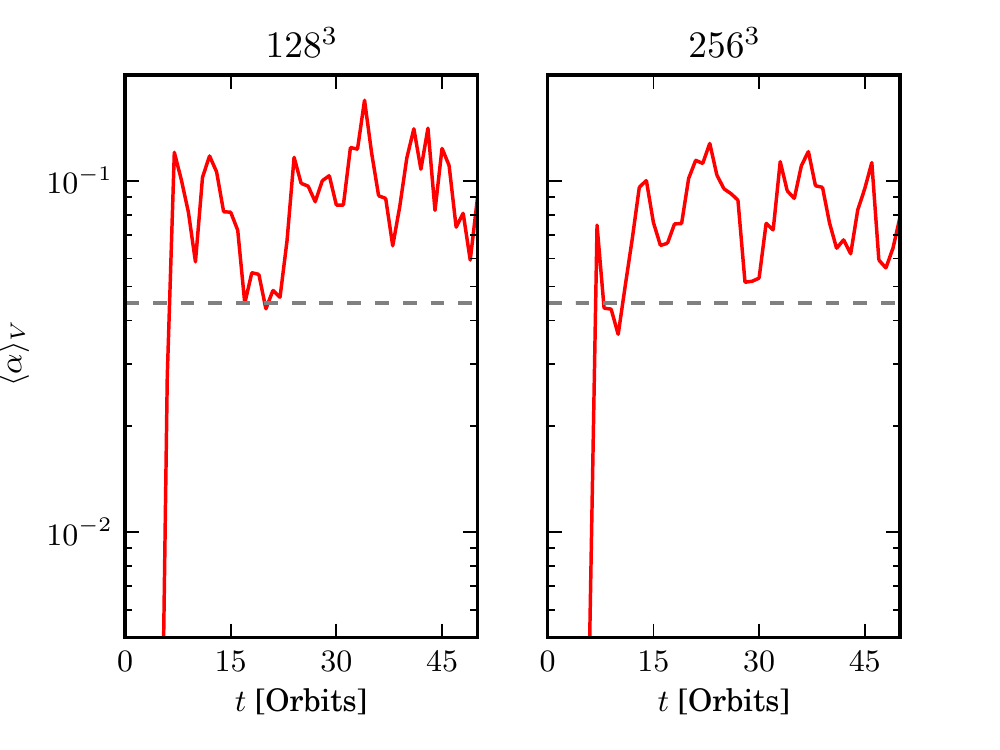}
\end{center}\caption{
Time series of $\langle \alpha \rangle_V$ for the Athena runs.  Compare to \Fig{beta_alpha_volavg}, bottom panels, the dashed gray reference line is the same as shown there.
\label{aalpha_volavg} }
\end{figure}

\begin{figure}[t!]\begin{center}
\includegraphics[width=\columnwidth]{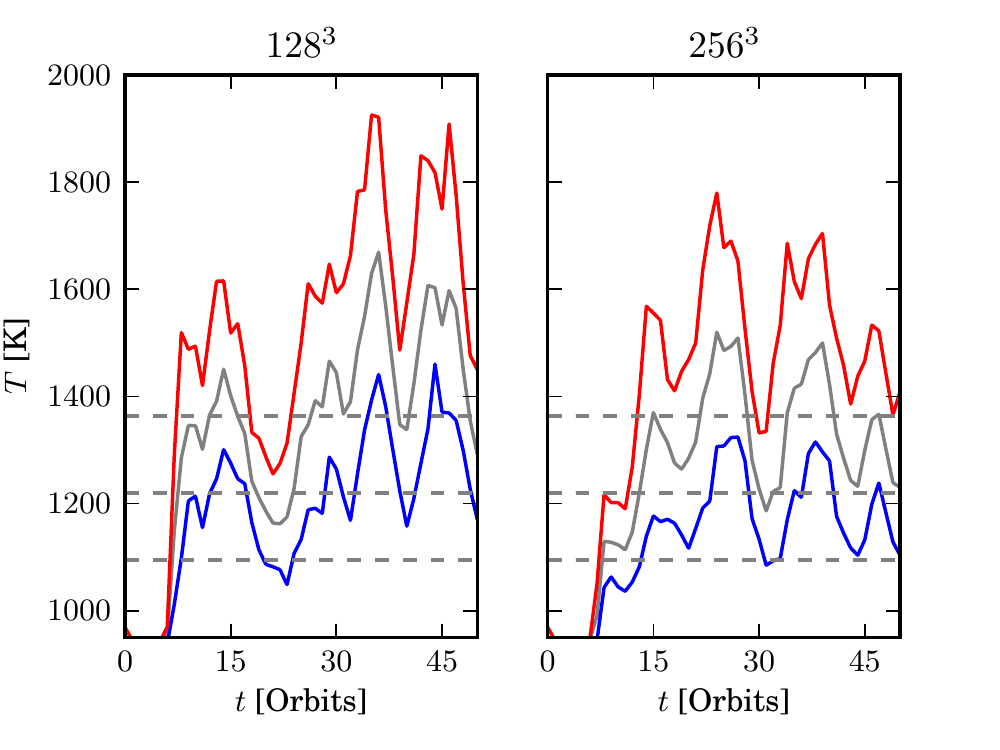}
\end{center}\caption{ 
$90$th percentile (red), $50$th percentile (gray) and $10$th percentile (blue) temperatures for the Athena runs.  
Compare to \Fig{res_check}, bottom panels, dashed gray lines are the same reference values shown there.
\label{athena_res_check} }
\end{figure}

To verify that the results from the Pencil Code resolution study are not dominated by any
possible local, nonconservative, energy dissipation, 
we have performed a smaller set of runs with Athena, which conserves energy to numerical accuracy.
These runs all start from $t=0$ and are run to $t=50$ orbits.
As expected, they display a different convergence behavior and
should be compared to the highest resolution Pencil Code result.
The volume average $\overline{\beta}$ at $128^3$ and $256^3$  shown in 
Figure~\ref{abeta_volavg} gives reasonable agreement with the Pencil
Code result, generally varying in the range $\overline{\beta} = $5--10.
The  $\langle \alpha \rangle_{\rm V}$ viscosity also agrees
reasonably, varying in the range 0.05--0.1, as shown Figure~\ref{aalpha_volavg}.
Finally, the variation in temperatures seen in Athena in Figure~\ref{athena_res_check} also agrees reasonably,
with the mean temperature typically $\sim 1200~\mathrm{K}$ and the spread between the
 $10\%$ and $90\%$ percentile values being typically $\sim 200~\mathrm{K}$.
 As the Godunov method employed 
 in Athena does not capture shocks through an explicit shock viscosity, it is difficult
 to decompose the heating rates as was done for the Pencil Code runs.
 Altogether, the Athena results are entirely consistent with the major
 result that current sheets can cause substantial local heating under
 the conditions we consider.

\section{Current Sheet Heating}
\label{sec_theory}

Figures~\ref{res_check} and~\ref{Slices_full} clearly show that there are
strong temperature variations in unstratified, net-vertical field MRI,
not only in time but also in space. These large temperature
fluctuations seem surprising given the modest overall $\overline{\beta}$
in the volume (\Fig{beta_alpha_volavg}).  The primary reason
for this surprising behavior is that our magnetic energy is
concentrated in large-scale azimuthal structures, which we call
magnetic field bundles. These interact to generate thin current sheets
with large horizontal extents that dissipate the magnetic energy.  In
this section we compute the temperature variations expected in this
situation, both for analytical understanding of our results, and to
extend the parameter space of their applicability.

\subsection{Adiabatic Compression}
The comparatively short perpendicular
length scale of these current sheets implies that as long as $\beta \ge 1$,
structures that are smaller than a scale height $H$ and survive for at
least an orbital timescale $\Omega^{-1}$ should be in near pressure
equilibrium with the bounding magnetic field bundles.
Accordingly, we can estimate the temperature variation by
assuming that adiabatic compression maintains pressure equilibrium
between the current sheet and the surrounding magnetic field bundles.
Outside of the current sheet, the pressure is altered by both
the growth of the magnetic field during the formation
of the bundles and adiabatic expansion into the current sheet,
while inside the current sheet the pressure is altered solely by
adiabatic compression.  

If we assume a sinusoidally varying magnetic
field, and the maximum value of the magnetic pressure is small enough
that we can expand linearly, then the adiabatic compression in the
current sheets is equal to the adiabatic expansion outside the current
sheets. Thus, if $\rho_0$, $T_0$ and $p_0$ are the density,
temperature and pressure in the disk before the generation of the
field bundles, the pressure $p_1$ of the system after
magnetic field growth and pressure equilibration
occurs will be 
\EQ 
p_1=p_0+\delta p_{\text{th}}=p_0-\delta p_{\text{th}} + \delta p_m 
\EN 
where $\delta p_m \simeq p_0/\beta_p$
is the maximum magnetic pressure increase, $\beta_p$ is the minimum plasma
$\beta$ associated with the peak magnetic field, and $\delta p_{\text{th}}$ is the amplitude of the thermal
pressure variation.  This implies that 
\EQ 
\delta p_{\text{th}}= \frac{p_0}{2 \beta_p}. 
\EN 

The temperature variation between the magnetized regions and
the current sheets is 
\EQ 
\frac{\delta T}{T_0} = \left(\frac{p_0+\delta p_{\text{th}}}{p_0-\delta
    p_{\text{th}}}\right)^{(\gamma -1)/\gamma} -1 \simeq
\frac{\gamma-1}{\gamma} \frac{1}{\beta_p}.
\label{deltaTadiab}
\EN
For our value of $\gamma=1.5$,
this reduces to $\delta T /T_0\simeq 1/(3 \beta_p)$.

\Eq{deltaTadiab} implies that strong magnetic fields will
always generate significant point-to-point temperature fluctuations through adiabatic compression alone,
but this effect alone clearly is insufficient to explain the simulated
behavior if $\beta_p \sim \overline{\beta} \sim 10$. However, there are
two further effects we must consider. Thermal diffusion will reduce
the temperature variation, but resistive heating in the current sheet
will increase it.

\subsection{Resistive Heating and Thermal Diffusion}
\label{OrbShear}
 
In this section, we analyze the time-dependent, resistive heating of current sheets
for magnetic fields whose peak magnetic field
is weak, with a minimum value of the magnetic to thermal pressure ratio $\beta_p^{-1} \ll 1$.
We expect that the azimuthal magnetic field bundles that bound the sheets
form from the
amplification of radial field variations by orbital shear. Such a
mechanism should
result in anticorrelation of radial and azimuthal magnetic fields. The
top panel of \Fig{line_cut} demonstrates that this is indeed the case
in our simulation.  

We can therefore approximate the system as an initial
radial field that varies only as a function of $z$ and decays
resistively.  
We do not include any variation in the $x$-direction as the 
current sheets are thin, extended structures.
As the azimuthal field bundles  
that appear in the simulations 
span the full box in azimuth, we also do not include any variation in the $y$-direction.
However, such $y$-direction variations could be an interesting study, 
as they would sharpen with the action of the background shear.
That radial field is sheared by a flow $u_y = Sx$ where
$S=-3\Omega_0/2$ is the Keplerian shear. The shear then generates an
azimuthal field that also decays resistively.  Under these conditions,
the induction \Eq{induction_2} becomes \citep{ 2005PhR...417....1B}:
\begin{align}
&\frac{\partial B_x}{\partial t} = \eta \frac{\partial^2 B_x}{\partial z^2}  \\
&\frac{\partial B_y}{\partial t} = -\frac{3\Omega_0}{2} B_x +\eta
\frac{\partial^2 B_y}{\partial z^2},
\end{align}
Assuming an initial radial field of $B_x=B_0 \sin(k z)$, and no
initial azimuthal field, the time evolution of the magnetic field is
\begin{align}
&B_x (t) = B_0 \exp({-t/\tau}) \sin(kz) \\
&B_y (t) = -\frac{3\Omega_0 t}{2} B_x =  -B_0 \left(\frac{3\Omega_0 t}{2}\right) \exp({-t/\tau}) \sin(kz),
\end{align}
where $\tau \equiv 1/(\eta k^2)$ is the resistive decay time associated with the wavenumber $k$.

The peak azimuthal field occurs at time $t=\tau$.  As long as the
ratio of the resistive timescale $\tau$ to the shear timescale $S^{-1}$,  namely $3\Omega_0 \tau/2 \gg
1$, $B_y$ dominates the magnetic field, with a peak strength of
\EQ
B_p = \frac{3\Omega_0 \tau}{2 \exp(1)} B_0. \label{B0}
\EN
In what follows, we consider only the heating due to this dominant
azimuthal field, neglecting $B_x$ for all 
purposes other than feeding $B_y$.

We normalize the peak azimuthal field strength to the initial thermal pressure
by defining the plasma-$\beta$ value for this peak azimuthal field with respect to the initial thermal pressure:
\EQ
\beta_p \equiv \frac{8 \pi p_0}{B_p^2}. \label{betam}
\EN
We assume that $\beta_p \gg 1$, so that we can linearize in $\beta_p^{-1}$.  In this limit, defining
$\beta_p$ with respect to $p_0$ rather than the thermal pressure at
the peaks of the perturbed magnetic field, where $\sin(kz)=\pm 1$, is equivalent to
expanding first order in $\beta_p^{-1}$ because the thermal pressure perturbation does contribute at lowest order.
We can use \Eqs{B0}{betam} to rewrite the initial seed field strength in terms of $\beta_p$:
\EQ
B_0^2 = \frac{32 \pi \exp(2) p_0}{9 \Omega_0^2 \tau^2 \beta_p}.
\EN
In Gaussian cgs units, the Ohmic resistive energy dissipation is
\EQ
 \frac{4 \pi \eta }{c^2} J^2 =  \frac{\eta}{4\pi} \left( \nabla \times \BB\right)^2 \, .
\EN
Following the approximation that the magnetic field is dominated by the $B_y$ component yields
\EQ
 \frac{4 \pi \eta }{c^2} J^2 =  \frac{2 \exp(2) p_0}{\beta_p \tau} \left(\frac{t}{\tau}\right)^2 \exp({-2t/\tau}) \cos(kz)^2.
\label{res_rate}
\EN
Note that in the case of Ohmic resistivity, 
the magnetic energy dissipation is exactly $90^{\circ}$ out of phase
spatially with the magnetic energy.  This is neither the case with
ambipolar diffusion (which scales with $\JJ \times \BB$), nor with numerical dissipation.

Between time $t=0$ and $t$, Ohmic resistivity deposits an energy
density 
\EQ
E = \int_0^t  dt' \frac{\exp(2) p_0}{\beta_p \tau^3}  t'^2 \exp({-2t'/\tau}) \left(1+\cos(2kz)\right) \label{eq_enerdep}
\EN
into the gas.
Using the relation for the thermal energy density $e=p/(\gamma-1)$, 
we can relate the total energy density $E$ deposited
resistively to the resulting temperature fluctuation to first order in $\beta_p^{-1}$: 
$\delta T = [(\gamma-1) E/p_0]~T_0$.

\subsubsection{Adiabatic Limit}
In the adiabatic limit, where the thermal diffusion coefficient $\mu \ll \eta$, and to first order in $\beta_p^{-1}$, there is
neither advective nor diffusive energy transport. 
Therefore the
highest point-to-point temperature variation occurs after all the
magnetic energy has resistively dissipated.  Then
\begin{eqnarray}
\delta T &=&  2 \int_0^{\infty} dt' \frac{(\gamma-1) T_0 \exp(2)}{\beta_p \tau^3}  t'^2 \exp({-2t'/\tau}) \\
&=& \frac{(\gamma-1) \exp(2) }{2\beta_p} T_0,  \qquad \frac{\mu}{\eta}\ll1
\end{eqnarray}
where the factor of  $\sin(\pi/2)-\sin(3\pi/2)=2$ comes from taking the difference between
the hottest and coldest points.

\subsubsection{Intermediate $\tau_E =\tau/2$ Case }
\label{sec_intermediate}
An analytical solution is possible in the particular case that
the thermal diffusion has a timescale $\tau_E = 1/(4\mu k^2) =\tau/2$, 
then Equation~(\ref{eq_enerdep}) gives:
\EQ
\frac{\partial \delta T}{\partial t} =\frac{2 (\gamma-1) T_0 \exp(2)}{\beta_p \tau^3}  t^2 \exp({-2t/\tau}) 
-\frac{\delta T}{\tau_E}.
\label{med_case_0}
\EN
In computing this solution note that the temperature fluctuation has wavenumber $2k$ and
\Eq{med_case_0} can be solved analytically, becoming
\EQ
\delta T = \frac{ (\gamma-1) \exp(2) t^3 \exp({-t/\tau_E})}{12 \tau_E^3 \beta_p} T_0,
\EN
which has a maximum for $t=3\tau_E = 1.5 \tau$ of
\EQ
\delta T =  \frac{9(\gamma-1)}{4 \exp(1) \beta_p} T_0.
\label{tT_med}
\EN
This underestimates the final result because we have ignored the adiabatic expansion
of the magnetic field bundle due to the magnetic pressure, which also generates temperature variations that
are first order in $\beta_p^{-1}$ and are spatially in phase with the resistive heating.  
However, that signal would have modestly diffused away by $t= 3 \tau_E$.

\subsubsection{Fast Cooling Limit }
Finally, if the temperature diffusion is very fast, the largest temperature variation will
occur at the time of fastest heating, or $t=\tau$, when $B=B_p$, and the thermal energy perturbation will be
just $(4 \pi/c^2) \eta J^2 \tau_E$.  
This means
\begin{eqnarray}
\delta T &=& \frac{2 (\gamma -1) }{\beta_p }\frac{\tau_E}{\tau} T_0 \\
 &=& \frac{(\gamma-1)}{2\beta_p} \frac{\eta}{\mu} T_0, \qquad \frac{\mu}{\eta}\gg1.
\label{tT_fast}
\end{eqnarray}
Here the heating 
simply scales by the ratio of magnetic
diffusion to thermal diffusion $\eta/\mu$.

\subsection{Comparison to Simulations}

In our simulation, we use temperature independent values of the
resistivity $\eta \sim 9 \times 10^{14}\ \mathrm{cm^2\ s^{-1}}$
(Table~\ref{table_parameters}), 
and the thermal diffusivity $ \mu \sim 1.8
\times 10^{14}\ \mathrm{cm^2\ s^{-1}}$ (Equation~(\ref{therm_diff_eff})), so
\begin{align}
& \tau_E k^2 = \frac{1}{4 \mu} = 1.4\times 10^{-15} \mathrm{\ s\ cm^{-2}}, \\
&\tau k^2 = \frac{1}{\eta} =  1.1 \times 10^{-15} \mathrm{\ s\ cm^{-2}}.
\end{align}
This means that, for any $k$, the timescale $\tau_E \sim \tau$, so we have slightly slower cooling than the regime of
\Eq{tT_med},  even though we are not in
the $\beta_p^{-1} \ll 1$ regime.
However, if we do apply \Eq{tT_med}  to our simulations
where $\beta_p \approx 1$ in both geometries, then the predicted $\Delta T/T_0=0.42$.
This agrees reasonably well with the results obtained in Figures~\ref{res_check} and \ref{res_check_run8}, 
given the differing $\beta_p$ regimes, especially considering that, as discussed in 
Section~\ref{sec_intermediate}, \Eq{tT_med} neglects the heating in the current 
sheet due to the adiabatic expansion of the magnetic field bundles.  This latter will
be particularly important when the system enters the  $\beta_p \approx 1$ regime during
particularly energetic phases.
Due to this, one would expect that  \Eq{tT_med}  underestimates the heating 
which occurs in our simulations, as is the case.

\section{Dissipation coefficients}
\label{opacity}

We have shown that the ratio of the resistivity to the thermal
diffusion plays an important role in limiting the point-to-point
temperature differences in protoplanetary disks. 
The MRI-criticality condition is $\Lambda_{0} \sim 0.1$, so in a disk
with $c_s \propto R^{-1/4}$ (appropriate for the minimum mass solar
nebula of \citealt{Hayashi81}, approximately the scaling of a passive irradiated disk,
and reasonable for a constant-$\alpha$ disk),
the critical resistivity scales with radius as
\EQ
\eta_{\text{crit}} \propto \beta_{\text{init}}^{-1} H^2 \Omega \propto R^1,
\label{eta_crit_sec_6}
\EN
where we have assumed a constant initial $\beta$ for the seed field.
Radiative thermal diffusivity scales as
\EQ
\mu \propto \frac{T^3}{\rho^2} \propto R^3
\label{mu_sec_6}
\EN
where we have assumed that the surface density scales with $\Sigma
\propto R^{-1}$ of a constant-$\alpha$ disk with the given temperature scaling \citep{Dullemond07},
and we have neglected the sublimation
of dust grains.  Note that letting the temperature vary with radius independently of $\eta$ implicitly assumes non-thermal ionization.

We chose our parameters to put the outer edge of the inner MRI active zone
at our location of $1$~AU.
Given the radial scalings above,
we can see that in the inner disk the
resistivity at the outer edge of the thermally-ionized MRI active zone
will usually dominate over thermal diffusion when the temperature is
just adequate to ionize the gas sufficiently to the critical value for
the MRI.  Further, the difference between the power-laws in
\Eqs{eta_crit_sec_6}{mu_sec_6} is
large enough that this conclusion is robust against modest changes to the disk model.  
However, that analysis becomes unimportant once the
temperature is high enough for the short--circuit instability to start,
as it causes resistivity and opacity to depend strongly on
temperature.

Beyond the dead zone, in the outer MRI-active regions of a
protoplanetary disk, the non-thermal ionization relies on a low
absorption column for cosmic rays.  This implies a lower
surface density, and hence a faster thermal diffusion than the inner
MRI-active region.  Therefore, when considering only the constant
Ohmic resistivity considered in this paper, the temperature variations will be
much smaller than those seen in the inner, thermally ionized
MRI-active region of the disk, following \Eq{tT_fast}.  If the
ionization has a strong temperature dependence, as occurs at the
metal lines suggested by \citet{2013ApJ...765..114D}, where ionization is
dominantly controlled by the adsorption of ions onto grain surfaces,
then the short--circuit instability of \citet{2012ApJ...761...58H} may
modify the behavior of current sheets, leading to larger heating.

\section{Discussion}
\label{conclusions}

We have shown that, under reasonable parameters, the MRI can generate
strong temperature variations across small regions. This has broad
implications, from the dynamics of the MRI itself through changes in
the gas density, through more indirect effects on the MRI such as the
short--circuit instability, to effects that go beyond the MRI entirely,
such as thermally processing solid material in protoplanetary disks.
It is clear from our results that even minimal MRI activity can
produce adequate temperature variations to trigger the short-circuit
instability in regions with temperatures close to the thermal ionization regime.

\subsection{Caveats}
The generalizability of this result beyond the regime of our
unstratified, net vertical field simulation does remain to be
confirmed.  In particular, including stratification could allow export
of magnetic energy vertically through buoyancy; while net-vertical
field MRI is relatively violent, so that relaxing these constraints
might lead to weaker current sheets, providing less heating.

We have further demonstrated that resolving the current sheet
structure of the MRI is a major undertaking.  Even though we plausibly
resolved the time-and-volume averaged magnetic field strength and
stresses with only $128^3$ simulations (about $26$ zones/scale
height), the heating terms were only beginning to be resolved at
$256^3$ (or about $50$ zones/scale height).  Further, this resolution
requirement was for nearly the highest  resistivity that would still
allow the MRI with the fastest growing wavelength contained in the 
thin aspect ratio box height.  
Lower levels of resistivity will require even higher
resolutions.

We employed a unstratified shearing box model, which limits the manner in which the 
large scale cooling of the disk can be taken into account.
Within this framework, we chose the fastest consistent value for the cooling timescale.
A vertically global model with full radiative transfer may have 
effectively slower thermal relaxation at the midplane, and therefore the
temperature fluctuations produced in our model may be underestimates.

A significant difference between the case treated in our simulations and an actual
protoplanetary disk is that 
we allow neither $\eta$ nor $\kappa$ to
vary with temperature.  
Accordingly, because our simulations already show strong temperature fluctuations, we can
state that any MRI simulations with temperature-independent resistivity 
and opacity in the temperature range we consider
are not self-consistent.  Our one-dimensional results on the short circuit
instability \citep{2013ApJ...767L...2M} suggest that substantially stronger
fluctuations would occur if a temperature-dependent resistivity and
opacity are included.

For our simulations to progress, we must choose initial conditions that are MRI unstable.
Awkwardly, as the MRI
develops it heats the box.  Were we to combine physically motivated, temperature-dependent resistivity
with an initial condition that was marginally unstable to the MRI, the rising temperatures and falling resistivities
would take the MRI away from marginal instability.
Indeed, initially MRI-stable regions have been seen to become 
unstable as the turbulent disk heats \citep{2012MNRAS.424.1977L,2014A&A...564A..22F}.
Such radial variations cannot be captured by a local shearing box model, 
so the method used in this paper has a limited ability to treat the 
dynamically evolving region 
at the potentially migrating, certainly self-determining inner edge of the dead zone.

\subsection{Consequences}

Our result further supports the hypothesis that magnetic dissipation
in general, and the short-circuit instability in particular may be
responsible for the deduced thermal histories of calcium--aluminum rich
inclusions (CAIs) and chondrules, as suggested by
\citet{2012ApJ...761...58H} and \citet{2013ApJ...767L...2M}.  

Several classes of igneous CAIs, particularly compact Type~A, Type~B,
and Type~C CAIs, appear to have been briefly heated back over their
melting temperatures at some time after their initial condensation
\citep{1986GeCoA..50.1785S,2005ASPC..341...15S}.  Being formed of the
most refractory minerals, CAIs may trace the inner, warm, and hence
MRI-active, regions of the protoplanetary disk where temperature
fluctuations generated by current sheets are unavoidable.
Chondrules are made of less refractory material than CAIs, and so
their thermal histories are often less extreme than those of
remelted CAIs \citep[e.g.][]{2006mess.book..253E}.  However, they still require heating to temperatures
where thermal ionization is easily capable of allowing the MRI to act.

Even in regions where
the MRI is active but where the ionization does not sharply increase 
with temperature
as required for the short-circuit instability,  the temperature 
fluctuations we have found could provide a route to annealing
silicates at more modest temperatures than are required to
melt chondrules or CAIs
\citep{2009ApJS..182..477S}.

The scale of the temperature variations seen suggests that significant
hysteresis may be possible in MRI activity. Once MRI begins, it may be
able to provide the ionization to sustain itself even if the
equilibrium disk becomes too neutral for it to continue 
\citep{2012MNRAS.424.1977L,2014A&A...564A..22F}.
The strongly
time varying accretion rates of protoplanetary disks, for example as
seen in FU Orionis type events \citep{Hartmann96}, predict, and can be
explained by, the inner active zone extending radially outwards
\citep{Armitage01}.  If MRI activity is adequate to generate local hot
regions with low enough resistivity to support the MRI, the MRI might
be active even in regions where the median resistivity is too high to
allow MRI.

This is a similar mechanism to that suggested by
\citet{2005ApJ...628L.155I}, although we favor current-driven heating
resulting in thermal ionization rather than relying on direct
non-thermal ionization by electrons accelerated in the current sheets.
However, the end result, that current sheets may have a significantly
lower resistivity, was explored by \citet{2012ApJ...760...56M}
\citep[see also erratum][]{2013ApJ...771..138M} who found that some
regions of what would otherwise be the dead zone can maintain an MRI
active state by such a mechanism.  Unlike the electron acceleration
mechanism of \citet{2005ApJ...628L.155I}, resistively heated current
sheets can be expected to be most effective at maintaining MRI when
the disk is massive, as this regime has the lowest thermal diffusion.

Another consequence of temperature variations is the broadening of ice
lines.  If the background temperature varies with radius as
$R^{-1/2}$, then $20\%$ temperature variations will broaden the
ice-line to $\sim 40\%$ of its ``mean'' position.  A significant
amount of material would then be repeatedly evaporated and recondensed
\citep{2013A&A...552A.137R}.  This effect is significant even with
quite small temperature variations: $4\%$ temperature variations will
still leave $8\%$ broadening, producing an annulus with a width of a few local scale heights.
The accretion crossing timescale for such an annulus exceeds
$\alpha^{-1}$, long enough for material within it to be processed
and reprocessed.  Multiple rounds of evaporation and condensation are unlikely to result in
the same grain porosity, and hence collision resilience, as direct
collisional growth, so larger grain sizes could be reached than those 
expected from collisional growth, as computed, for example, by \citet{2012A&A...544L..16W}.

\acknowledgements
We thank D.~Ebel for useful discussions, and G.~Lesur for a
constructive referee report that helped generalize our results.
Supercomputing resources at J{\"u}lich Supercomputing Centre (PRACE project ``Protostars: From Molecular Clouds to Disc Microphysics'') and at DeIC/KU in Copenhagen are gratefully acknowledged.
The research leading to these results has received funding from the 
People Programme (Marie Curie Actions) of the European Union's Seventh Framework Programme 
(FP7/2007-2013) under REA grant agreement 327995 (C.P.M.), and the
U.S. National Science Foundation under CDI grant AST08-35734 and AAG grant AST10-09802 (A.H.
and M.-M.M.L.).

\bibliographystyle{apj}

\end{document}